\documentclass{JHEP3}
\input epsf
\usepackage{amsmath}
\usepackage{amssymb}

\newcommand{\half}{\frac12}

\newcommand{\refeq}[1]{Eq.~(\ref{eq:#1})}

\newcommand{\beq}{\begin{eqnarray}}
\newcommand{\eeq}{\end{eqnarray}}

\def\tr{{\rm\ Tr}}

\def\mn{{\mu\nu}}

\newcommand{\vev}[1]{\langle#1\rangle}
\def\be{\begin{equation}}
\newcommand{\bel}[1]{\be\label{#1}}
\def\ee{\end{equation}}
\newcommand{\eref}[1]{(\ref{#1})}
\newcommand{\Eref}[1]{Eq.~(\ref{#1})}
\newcommand{\rem}[1]{}
\def\tr{{\rm tr}}
\def\half{{1\over 2}}
\def\NN{{\cal N}}

\def\elll{\ell}

\def\none{$\NN=1$}
\def\nonestar{$\NN=1^*$}
\def\ntwo{$\NN=2$}

\def\nfour{$\NN=4$}

\newcommand{\FF}[3]{F^{(#1)#2}_{#3}}

\newcommand{\tFF}[3]{\tilde F^{(#1)#2}_{#3}}

\def\QQ{{\cal Q}}

\def\bit{\begin{itemize}}
\def\eit{\end{itemize}}

\def \be {\begin{equation}}
\def \ee {\end{equation}}
\def \bea {\begin{eqnarray}}
\def \eea {\end{eqnarray}}
\def \half{\frac{1}{2}}
\newcommand{\ket}[1]{{\left | {#1}\right \rangle}}
\newcommand{\bra}[1]{{\left \langle {#1}\right |}}
\newcommand{\tmmathbf}[1]{\boldsymbol{#1}}

\newcommand{\q}{\bar{q}}
\newcommand{\m}{\bar{m}}
\def\xp{{\bar{x}_\perp}}
\def\xp{{\bf{x}}}

\def\rflav{r_f}
\def\rfour{r_1}
\def\rgrav{r_2}

\title{Quarkonium from the Fifth Dimension}

\author{Sungho Hong$^{ab}$,Sukjin Yoon$^{a}$,
and Matthew J. Strassler$^{a}$\\
$^{a}$Department of Physics and Astronomy\\
P.O Box 351560, University of Washington\\
Seattle, WA 98195\\
\\
$^{b}$Department of Physics and Astronomy\\
University of Pennsylvania\\
Philadelphia, PA 19104-6396}

\abstract{Adding fundamental matter of mass $m_Q$ to \nfour\ Yang Mills theory,
we study quarkonium, and ``generalized quarkonium'' containing light
adjoint particles.  At large 't Hooft coupling the states of spin
$\leq 1$ are anomalously light (Kruczenski {\it et al.},
hep-th/0304032).  We examine their form factors, and show these
hadrons are unlike any known in QCD.  By a traditional yardstick they
appear infinite in size (as with strings in flat space) but we show
that this is a failure of the yardstick.  All of the
hadrons are actually of finite size $\sim\sqrt{g^2N}/m_Q$, regardless
of their radial excitation level and of how many valence adjoint
particles they contain.  Certain form factors for spin-1 quarkonia
vanish in the large-$g^2N$ limit; thus these hadrons resemble neither
the observed $J/\Psi$ quarkonium states nor
$\rho$ mesons.}


\preprint{hep-th/0312071\\
UW/PT 03-31\\
UPR-1057-T}

\begin{document}

\section{Introduction}

The discovery that gauge theory at large 't Hooft coupling can be
described using string theory \cite{Maldacena} has given rise to many
interesting developments in both fields.  The original idea has been
extended to purely four-dimensional confining gauge theories
\cite{Terningetal,nonestar,KlebStras}.  These examples provide the
first toy models of QCD whose kinematics is that of the real world.
In particular, these theories share with QCD the property of
four-dimensional approximate scale invariance in the ultraviolet, as
well as confinement in the infrared.  Consequently, the differences
between these models and QCD are largely due to dynamical, rather than
kinematical, issues.  Indeed, these theories smoothly become QCD-like
when the 't Hooft coupling is small.

However, most of the theories studied up to now differ from QCD in a
significant way: their matter comes in representations with of order
$N^2$ fields, and they are all neutral under some portion of the
center of the gauge group.  These differences from QCD are much more
important, both kinematically and dynamically, than the fact that many
of these theories are supersymmetric.  The reasons for this are clear.
Quarks in the ${\bf N}$ representation of $SU(N)$ color introduce new
features into the $1/N$ expansion; quark pair-production makes
confining flux tubes unstable; baryons now appear in the spectrum; and
there are new flavor symmetries which may or may not be explicitly or
spontaneously broken.  Thus the theory without quarks in the ${\bf N}$
representation, supersymmetric or not, differs signficantly from the
theory which contains them.  It is thus important to the development
of these toy models of QCD to introduce matter in the fundamental
representation.  In the limit where the number of flavors $N_f$ is
much less than $N$, this was considered in \cite{Aharony,Grana,Naculich};
a simpler method was invented in \cite{KK} and studied
further in \cite{KKW, Myers}.  Still more recently, there has been
additional work in other related contexts \cite{Sakai}.

One of the obvious objects to study in a theory with quarks is
quarkonium.  At small 't Hooft coupling this is just a hydrogenic
atom, but at large 't Hooft coupling the system is highly
relativistic.  The quarkonium spectrum at large 't Hooft coupling, in
the model studied in \cite{Myers}, is remarkable from the
field-theoretic point of view.  Its details could not have been
guessed from any known theoretical argument or from any aspect of
the observed QCD spectrum, and it is profoundly tied up with the
representation of the four-dimensional gauge theory as a
higher-dimensional string theory.  Still more puzzles emerge in states
built from a quark, an antiquark, and one or more particles of the
much lighter adjoint matter.  These states would naively be expected
to be qualitatively different in size and structure from pure
quarkonium. Instead, it has been found \cite{Myers} that these states
are quite similar to pure quarkonium at large 't Hooft coupling.

In order to gain better insight into the structure and couplings
of these bound states, we have computed some of their form
factors and transition matrix elements with respect to various
conserved currents.  Fourier transforming the form factors, we find
that all the hadrons are more or less the same size --- larger than
one would expect for quarkonium, and smaller than one would expect
when adjoint matter is bound to the quark-antiquark system.  Our
results do not solve any mysteries, but they do raise interesting
questions and suggest other calculations to do in future. We are also
led to a conjecture about the substructure of the hadrons.

We review the results of \cite{KK} and \cite{Myers} in the
following section.  Sections 3, 4 and 5 contain our methodology,
a summary of our results, and the detailed computations.
The physical implications of our results are discussed in
Section 6.

\section{Preliminaries}

The introduction of matter in the fundamental representation into
theories with gravitational dual descriptions has been considered by a
number of authors \cite{Aharony,Grana,Naculich}.  However, theoretical
prejudices about the appropriate systems, and technical difficulties
with those that were investigated, delayed progress for some time.
Recently, Karch and Katz \cite{KK} cut the Gordion knot, pointing out
that many interesting questions could be addressed in the simplest
possible brane construction with fundamental matter: a small number
$N_f$ of D7 branes in the vicinity of a large number $N$ of D3 branes.

\subsection{The theory in question}

The field theory corresponding to this arrangement of branes consists
of \nfour\ $SU(N)$ Yang-Mills coupled in an \ntwo\ supersymmetric
fashion to $N_f$ hypermultiplets in the fundamental representation.
We will write the \nfour\ vector multiplet as an \none\ vector
multiplet $W_\alpha$ and three chiral multiplets
$\Phi_1,\Phi_2,\Phi_3$ in the adjoint represntation.  The
hypermultiplets are given as \none\ chiral multiplets $Q^r,\tilde Q_r$
($r=1,\dots, N_f$) in the fundamental and antifundamental
representation respectively, and we will call their scalars
``squarks'' and ``antisquarks'' and their fermions ``quarks'' and
``antiquarks.''  Written in \none\ language, the theory consists of
kinetic terms for all the fields (along with \none\ superpartners of
the kinetic terms) and a superpotential of the form
$$
W= \sqrt{2}\ {\rm tr}\Big([\Phi_1,\Phi_2]\Phi_3\Big) + \sum_{r=1}^{N_f}
Q^r\Phi_3\tilde Q_r + m_r Q^r\tilde Q_r
$$
where $m_r$ is the mass of hypermultiplet $r$ and the trace is
over color indices.

The theory has an $SO(4)\approx SU(2)\times SU(2)$ symmetry,
consisting of an $SU(2)_\Phi$
symmetry rotating $\Phi_1$ and $\Phi_2$ and an $SU(2)_R$ \ntwo\ R-symmetry.
The charges of the fields under
these symmetries are shown in the table, where we write
$\Phi_1=X^4+iX^5$, $\Phi_2=X^6+iX^7$, and $\Phi_3=X^8+iX^9$, the
adjoint fermions as $\lambda_1$, $\lambda_2$, $\lambda_3$,
$\lambda_4$, and the quarks as $\psi$, $\tilde\psi$.

\

\

\begin{tabular}{|c|c|c|c|}
  \hline
   & $SU(2)_{\Phi} \times SU(2)_{\cal R}$ & $U(1)_R$ & $U(N_f)$ \\
  \hline
  $X^{4,\cdots, 7}$ & $(\half,\half)$ & 0 & 1 \\
  $X^{8,9}$ & $(0,0)$ & $\pm 2$ & 1 \\
  $\lambda_1,\lambda_2$ & $(\half,0)$ & $-1$ & 1 \\
  $\lambda_3, \lambda_4$ & $(0,\half)$ & $+1$ & 1 \\
  $Q, \tilde{Q}^\dag$ & $(0,{\half})$ & 0 & $N_f$ \\
  $\psi, \tilde{\psi}^\dag$ & $(0,0)$ & $\mp 1$ & $N_f$ \\
  \hline
\end{tabular}

\

\

\noindent The masses $m_r$ are the eigenvalues
of a matrix which transforms in the adjoint of $U(N_f)$.
If all masses $m_r$ are
zero,
the
$U(1)_R$ \none\ R-symmetry is preserved by the
superpotential.  Nonzero masses break this $U(1)$ R-symmetry, but
leave invariant the $SO(4)$.  The $U(N_f)$ is also generally broken by
the masses, but if all the masses are equal, as we will assume
throughout, $U(N_f)$ is preserved.

The string dual description of this field theory \cite{KK}, for large
't Hooft coupling,\footnote{We use the following notation: $g_{YM}$ is
the \nfour\ gauge coupling, $\alpha \equiv g_{YM}^2/4\pi$,
$\lambda\equiv g^2
N$ is the 't Hooft coupling, $g_s$ is the type IIB closed string
coupling, and the couplings are related under gauge/string duality by
$g_s=\alpha$.}  is simply given as $N_f$ D7 branes placed as probes
inside $AdS_5\times S^5$.  Each D7 brane introduces a hypermultiplet
of mass $m$, where $m$ is proportional to the minimum value of the
$AdS_5$ radial coordinate $r$ to which the D7 brane descends.  At any
larger and fixed value of $r$, the D7 brane fills out an $S^3$
subspace of the $S^5$.  We will give more details on this brane
construction below.

 The theory's reduced supersymmetry is of little concern, but it does
seem at first to have a serious dynamical problem.  The gauge coupling
now has a positive beta function, making the ultraviolet definition of
the theory problematic, and potentially destroying any hope of making
sense of the theory using gauge-string duality.  In particular, the
dilaton is no longer constant.  Where the dilaton (equivalently, the
gauge coupling of the gauge theory) becomes of order 1, one ought to
do an S-duality transformation; but this would turn the D7-branes into
magnetic 7-branes, which are very difficult to handle.

However, the gauge beta function is very small.  In particular, the
beta function for $g_{YM}$ is of order $g_{YM}^3 N_f$, so the beta
function for the 't Hooft coupling $\lambda$ is of order
$\lambda^2 N_f/N$.  Another way to say this is that at large $N$ and
small $N_f$ this is a naturally ``quenched'' theory: all quantum
effects due to the fundamental matter are parametrically suppressed by
$N_f/N$.  The squarks and quarks are simply probes of the dynamics of
the \nfour\ sector.  From the supergravity point of view, this
suppression is essentially the statement that the backreaction,
through the variation of the dilaton, metric and other fields, caused
by the D7-branes is negligible out to exponentially large $AdS$
radius.  Consequently, as Karch and Katz argued, we can study to
leading order in $N_f/N$ all of the issues that can normally be
investigated in quenched QCD, including the spectrum of hadrons.

\subsection{The (s)quark-anti(s)quark hadrons}

Karch, Katz and Weiner \cite{KKW} considered the formation of
heavy-light mesons in this theory, working at large 't Hooft coupling
$\lambda\gg 1$.  They found a number of surprising
results, including confinement of quarks without formation of flux
tubes (what one might call ``Gribov confinement.'' \cite{gribov})

Another natural issue for study is quarkonium, namely bound states of
a quark and antiquark of equal mass.  At weak coupling ({\it i.e.},
small $\lambda$) this is simply the problem of the hydrogen atom, but at
strong coupling we do not {\it a priori} know what to expect.

A seemingly different problem is that of a quark and antiquark bound
to one or more of the $\Phi_i$ particles.  In the simplest D3-D7
system, these particles are massless.  An easier system to think about
is obtained by giving the $\Phi_i$ small bare masses, so that (as in
\nonestar \cite{nonestar}) confinement of flux occurs at a distance
very long compared to the quarkonium mass and length scales.
Then we can imagine stable bound states of a quark,
antiquark and some number of the much lighter $\Phi_i$ particles.
This problem has not been much studied at small 't Hooft coupling, but
one would expect these bound states, in which the quark and antiquark
combined must be in the adjoint representation, would be very
different in size and shape from the simple quarkonium bound states.

Surprisingly, at large 't Hooft coupling both problems can be
addressed simultaneously, and on the same footing.
This was done by  Kruczenski, Mateos, Myers and Winters
\cite{Myers}.  They derived the spectrum and higher-dimensional wave
functions of mesons consisting of one (s)quark and one anti(s)quark of
equal mass, along with some number of massless adjoint particles.  The
spectrum for states of large spin ($s\gg \sqrt{\lambda}$) was found to be
the same as that of hydrogen; this is to be expected, since in this
limit the (s)quarks move nonrelativistically even though the coupling
is strong.  For states of lower spin it was shown that the spectrum
has a Regge-like relation betweeen mass and spin: $m^2\propto s$.  And
states of spin $1$, $\half$ and 0 were found to be extraordinarily
deeply bound --- so deep that the mass of the hadron is a tiny
fraction of the mass of its (s)quark constituents.  It is these
anomalously light states of low spin --- states that are described in
the string theory variables by modes of massless higher-dimensional
fields --- that are the focus of this article.  We now review their
properties in detail.

We generally follow the setup in~\cite{KK}.  The
near-horizon geometry of $N$ D3-branes filling the 0123 directions of
the ten-dimensional space, and placed at the origin of the 456789
coordinates, is given by $AdS_5\times S^5$.  We write the metric on
the Poincare' patch variously as\footnote{We use $-+++$ signature.}
$$
ds^2 = {r^2\over R^2} (dx^\mu)^2 + {R^2\over r^2} dr^2 + R^2 d \Omega_5^2
= {r^2\over R^2} \eta_{\mn}dx^\mu dx^\nu + \sum_{c=4}^9 {R^2\over r^2} (dx^c)^2
$$
where $c=4,5,6,7,8,9$, $\mu=0,1,2,3$, $r^2\equiv{\sum_c (x^c)^2}$,
and $R^2=\alpha' \sqrt{4\pi g_s N}$.
Note $r=0$ is the horizon of $AdS$ and $r=\infty$ is its boundary.
The D7-branes fill the 01234567 directions; each is placed
at some position $x^8+ix^9= m_Q\alpha'$.  We will make all the
masses equal in this paper, so without loss of generality we
can take the masses to be real ($x^8=m_Q\alpha'\equiv L$, $x^9=0$.)
The induced metric on the D7-branes is then
\bea
d s^2 &=&  {r^2\over R^2} \eta_{\mn}dx^\mu dx^\nu + \sum_{c=4}^7
{R^2\over r^2} (dx^c)^2 \cr
&=&
\frac{L^2}{R^2}(\varrho^2 + 1) \eta_\mn dx^\mu dx^\nu +
\frac{R^2}{\varrho^2 + 1} d \varrho^2 +R^2
\frac{\varrho^2}{\varrho^2 + 1} d \Omega_3^2 \ ,
\eea
where $\varrho^2= \frac{r^2}{L^2}-1$, and the $S^3$ involves the
angular coordinates in the four-dimensional space spanned by
$x^4, x^5,x^6,x^7$.

We will use various coordinates for
our calculations which we summarize here:
\bel{coords} v = (L/r)^2 \ ,  \ w = 1-v\  ;\ \varrho^2 = v^{-1}-1
= \frac{w}{1-w} \ . \ee
Note the boundary of the space is located at $r=\infty$, $\rho=\infty$,
$v=0$ and $w=1$, whereas the point on the D7-brane which is closest
to the $AdS$ horizon at $r=0$ is at $r=L$, $\rho=0$, $v=1$ and $w=0$.

It is useful to define an overall
mass scale
$$
m_h \equiv
L/R^2= m_Q / \sqrt{4\pi g_s N} = m_Q/\sqrt{\lambda}\ .
$$
Barred quantities will be defined relative to $m_h$ --- for instance,
we will often use dimensionless momenta
$\q\equiv \sqrt{|q^2|}/m_h$ and dimensionless masses $\m = m/m_h$.

There are two real scalar fields and one gauge field on the D7 brane
worldvolume; theses are the massless modes of the open strings whose ends
are on the D7-branes.\footnote{Note these 7-7 strings are not dynamical
in the gauge theory!  In the Maldacena limit of the D3-D7 system, the 3-3
strings (the \nfour\ sector) and the 3-7 strings (the hypermultiplets)
are dynamical, whereas the closed strings and 7-7 strings act as background
fields.  In the dual string description, the 3-3 and 3-7 strings are absent,
and the closed and 7-7 strings are the dynamical degrees of freedom.}
 If there are $N_f$ D7-branes, there are $N_f^2$ such
fields, transforming in the adjoint representation of $U ( N_f )$. The
modes of the gauge field, when reduced to five dimensions, break up
into spin-one and spin-zero modes in five dimensions.  The spin-one
modes are associated with the $U(N_f)$ flavor current, and operators
built by adding $\Phi$ fields to the current:
\bel{eq:QphidQ}
 ( Q^{\dag} \Phi^\elll  Q - \tilde{Q} \Phi^\elll  \tilde{Q}^{\dag} )_{\theta
   \bar{\theta}} = Q^{\dag} \Phi^\elll  \partial^{\mu} Q + \psi^{\dag}_Q \Phi^\elll
   \sigma^{\mu} \psi_Q - \tilde{Q} \Phi^\elll  \partial^{\mu} \tilde{Q}^{\dag} +
   \cdots
\ee
where $\Phi^\elll $ stands for any product of $\Phi_1$ and $\Phi_2$
which is a symmetric and traceless representation under $SO(4)$. These
operators have conformal dimension $\Delta_1\equiv\elll + 3$ and
transform as $\left( \frac{\elll }{2}, \frac{\elll }{2} ) \right.$
under $SO(4)\approx SU(2)_\Phi\times SU(2)_R$, and as adjoint under
$U(N_f)$.

When these operators act on the vacuum they can create a spin-one
bound state of a $Q$, a $\tilde Q$, and $\elll $ $\Phi$ particles
(plus any number of gluons, gluino pairs, and $\Phi$--anti-$\Phi$
pairs, of course.)\footnote{They cannot, in this theory, create a
spin-zero state; this is not a general feature of the large 't Hooft
coupling limit, however.} In particular, for $\elll =0$ this is a
spin-one quarkonium state, much the same as the J/$\psi$ or the
$\Upsilon$ in QCD.  The bound state spectrum, and the bulk
wave functions of the individual bound states, were computed in
\cite{Myers}, where they were termed ``type $II$''.
These hadrons are modes of the 8-dimensional gauge
boson $A_M$, of the form
\bel{TypeII} A_{\rho} = 0,\quad A_{\alpha} = 0,\quad A_{\mu} =
\zeta_{\mu} \phi^{II}
( \rho ) e^{i k \cdot x}
   \mathcal{Y}^{\elll } ( S^3 ),\qquad k\cdot \zeta =0
\ee
\begin{eqnarray*}
\phi^{II}_{\elll ,n} & = & (C_{\elll n}^{II}/R^2) \varrho^{\elll
}(1 + \varrho^2)^{-1-n-\elll } F
(-n, -1-n-\elll ; \elll  + 2 ; - \varrho^2 )\\
& = & (C_{\elll n}^{II}/R^2) w^{\elll /2} (1-w)^{-\elll /2}
F(-1-n-\elll ,2+n+\elll  ;\elll +2;w)\\
& = & (\hat C_{\elll n}^{II}/R^2)
v^{(\elll+2) /2} ( 1 - v
)^{\elll  / 2} P_n^{(\elll  + 1,\elll+1)} ( 2 v - 1 )\ ,
\end{eqnarray*}
where $n\geq0$, $\elll \geq 0$,
$$C_{\elll n}^{II}= \sqrt{2(2n+2\elll +3)
{{n+2\elll +2}\choose{\elll +1}} {{n+\elll +1}\choose{\elll +1}}
}
= {{n+\elll+1}\choose{\elll+1}}
\hat
C_{\ell n}^{II} \ .
$$
Here $P_n^{( \alpha ,
\beta )} ( x )$ denotes a Jacobi polynomial. The masses of these
states are
$$M_{II}^2 = 4 m_h^2 (n+\elll +1)(n+\elll +2) \ ,
$$
where again $m_h\equiv L/R^2=  m_Q(\lambda)^{-1/2}$. By
coincidence the $\elll=0$ states also appear in the model of
\cite{SonSteph}.

Another mode which is easy to identify is the mode $I-$ in
\cite{Myers}, whose conformal dimension is $\Delta_0\equiv\elll + 1\
(\elll \geq 1)$, and which transforms as $\left( \frac{\elll-1}{2},
\frac{\elll+1}{2} \right)$ under $SO(4)\approx SU(2)_\Phi\times SU(2)_R$.  For
$\ell=1$ it has dimension 2 and is a triplet under $SU(2)_R$.  This
uniquely picks it out as a multiplet containing the \none\ chiral
operator
\bel{QphiQ} ( \tilde{Q} \Phi^{\elll-1} Q )_{\theta, \bar{\theta} = 0} = \tilde{Q} \Phi^{\elll-1} Q +
   \cdots
\ee
along with other modes related by $SO(4)$ (mainly given
by replacing $Q$ with $\tilde Q^\dagger$, $\Phi_1$ with $\Phi_2$ or
$\Phi_1^\dagger$, etc.)  The mode $I-$ is a mode of the 8-dimensional gauge
field with its component on the $S^3$ nonzero:
\bel{TypeI} A_{\mu} = 0,\quad A_{\rho} = 0,\quad A_{\alpha}\ =
\phi^{I-} ( \rho ) e^{\
i k \cdot x}
   \mathcal{Y}_{\alpha}^{\elll ,-} ( S^3 )
\ee
\begin{eqnarray*}
\phi^{I-}_{\elll ,n}  & = & (C^{I}_{\elll n}/L) \varrho^{\elll +1}
(1+\varrho^2
)^{-1-n-\elll } F(-n, 1-n-\elll  ;\elll +2 ; -\varrho^2)\\
& = &  (C^{I}_{\elll n}/L) w^{\frac{\elll +1}{2}} (1-w)^{-(\elll
-3)/{2}} F
(1-n-\elll , 2+n+\elll  ;\elll +2 ; w) \\
& = & (\hat C^{I}_{\elll n}/L)
v^{(\elll  + 1) / 2} ( 1 - v )^{( \elll  + 1 ) / 2}
   P_n^{( \elll  + 1, \elll  - 1 )} ( 2 v - 1 )
\end{eqnarray*}
where $n\geq 0$, $\elll \geq 1$, and
$$
C_{\elll n}^{I}= \sqrt{2(2n+2\elll+1) {{n+2\elll}\choose{\elll+1}}
{{n+\elll+1}\choose{\elll+1}} }
=
 {{n+\elll+1}\choose{\ell+1}}
\hat C_{\ell n}^I \ .
$$
These scalars have masses
$$M_{I-}^2 = 4 m_h^2 (n+\elll )(n+\elll +1) \ .
$$
Since we will not use the type $I+$ hadrons, we will henceforth
discard the minus sign in $I-$ when referring to these states.

These are the two classes of (s)quark-containing operators and states
that we will use for calculations in the rest of this paper.  We
will also consider matrix elements of the flavor current, which is
the operator appearing in \refeq{QphidQ} with $\ell=0$.  This will
require us to know the nonnormalizable mode of the corresponding
five-dimensional gauge boson at spacelike $q^2=\q^2 m_h^2$, which
is of the form
\bel{flavornon}
\phi^{II}_{non} (w;\alpha) = \frac{\pi\alpha(1+\alpha)}{\sin(\pi
\alpha)} F(-\alpha,1+\alpha ;2;w),
\ee
where $\alpha= \half\left(-1+\sqrt{1-\q^2}\right)$.

For completeness we briefly comment about the other classes of hadrons
appearing in \cite{Myers}.  There is a set of complex operators
$\psi_{\tilde{Q}} \Phi^{\ell-1} \psi_Q+\cdots$ which are obtained from
the operators \eref{QphiQ} by the action of two supersymmetry
generators; these complex modes create the two ``scalar modes'' in
\cite{Myers}.  The type $III$ and type $I+$ hadrons are harder to
identify, as there are a number of candidate operators, of which only
particular linear combinations are chiral.\footnote{Note there are no
chiral operators of protected dimension containing $\Phi_3$ and $Q$
together, because the form of the superpotential implies $\Phi_3Q$ and
$\tilde Q\Phi_3$ are simply proportional to $Q$ and $\tilde Q$,
respectively, within the chiral ring.}

\subsection{The \nfour\ sector}\label{sec:N4sector}

We will also consider certain modes of the
\nfour\ sector, which appear as  states of the ten-dimensional
supergravity on $AdS_5\times S^5$.  We will only
need the spin-one states created by acting on the vacuum with the
$SO(4)$ current $J_4^{a\mu} \sim {\rm tr}\ \Phi_i^\dagger \partial_\mu
\Phi_j (T^{a}_{SO(4)})_{ij} + \cdots$ and the spin-two states given
by acting on the vacuum with $T^{\mu\nu} \sim {\rm \tr} \
F^\mu_{\rho}F^{\rho \nu} +\Phi_i^\dagger \partial^\mu\partial^\nu \Phi
+ \cdots$.  (The traces are over color indices.)
However, it will be convenient to consider also a spin zero operators which
has a close relationship with the spin one and two operators. The spin zero
operator corresponds to the ten-dimensional metric element which is a singlet
under the transformation on the seven brane. Under the breaking $S O ( 6 )
\rightarrow S O ( 4 ) \times S O ( 2 )$, the traceless symmetric rank two
tensor representation $\tmmathbf{2} \tmmathbf{0}$ branches to
$(\tmmathbf{1},\tmmathbf{1}) \oplus (\tmmathbf{9},\tmmathbf{1}) \oplus
(\tmmathbf{1},\tmmathbf{2}) \oplus (\tmmathbf{4},\tmmathbf{2})$.  The
singlet
corresponds to the dimension-two $SO(4)$--singlet
operator ${\cal S}\sim \Phi_1^2 + \Phi_2^2 - 2\Phi_3^2$.

\subsubsection{The continuous spectrum of the conformal theory}

We will consider the operators whose spins range from zero to two. The
non-normalizable mode for the operator with spin $S$ is \cite{Modes}
\be\label{eq:ModeNNGeneral}
\left( \frac{L^2}{R^2 v} \right)^{S - 1} \frac{( \q \sqrt{v} )^S}{S!}
   K_S ( \q \sqrt{v} ).
\ee
where $\bar q =\sqrt{q^2}/m_h$ and $q^2>0$ is spacelike.  Note that
$L^2/R^2 v$ in the prefactor is the warp factor of the four
dimensional part of the metric. It is needed in order to have the
correct boundary behavior of the non-normalizable mode.

The $SO(4)$ current is associated under AdS/CFT duality with a
five-dimensional gauge boson $\hat A$, which descends from the
off-diagonal elements of the ten-dimensional metric.  In particular,
let $v^a$ be a Killing vector on the three-sphere
which points purely parallel to the D7-brane
world-volume (and thus leaves it invariant.)  Then the metric elements
$h_{\mu a} = \hat A_\mu v_a, h_{ra}=\hat A_r v_a$ (where $ds^2 =
ds^2_{AdS_5\times S^5}+ h_{\mu a} dx^\mu dx^a$)
define the $SO(4)$
gauge boson on $AdS_5$. In radial gauge $\hat A_r=0$, the
non-normalizable modes of $\hat A$
at spacelike $q^2$ are
\bel{TheModesAnn}
\hat A_\mu(q) =\epsilon_\mu \q \sqrt{v} K_1 ( \q \sqrt{v} ) \
\ee
which are normalized following~\eref{eq:ModeNNGeneral}
so that $\hat A_\mu \to \epsilon_\mu$ as
$v\to 0$. The corresponding normalizable
modes, at specific values of timelike $q^2 = -|q^2|\equiv - m_h^2 \bar m^2$,
are of the form
\bel{TheModesAn}
\hat A_\mu(q) \propto \epsilon_\mu \bar m \sqrt{v} J_1(\m \sqrt{v})\ .
\ee
One can build normalizable wave packets from the
normalizable modes, although,
just as with plane waves in flat space, there
is no normalized version of the modes in~\eref{TheModesAn}.

For the spin-two current we need the modes of the traceless
$AdS_5$ part of the
metric.  We consider
traceless fluctuations $ds^2 =
ds^2_{AdS_5\times S^5}+ h_{\mu \nu} dx^\mu dx^\nu$, $h_\mu^\mu=0$,
in radial gauge $h_{rr}=0, h_{r\mu}=0$.
The normalizable modes take the form
\bel{TheModesgn}
h_{\mu\nu}(q) \propto \epsilon_{\mu\nu} J_2(\m \sqrt{v})
\ee
while the nonnormalizable modes are
\bel{TheModesgnn}
h_{\mu \nu} ( q ) = \epsilon_{\mu\nu} \frac{L^2}{R^2 v} \cdot
\frac{ \q^2 v}{2} K_2 ( \q
\sqrt{v} )
\ .
\ee
As $v \rightarrow 0$, the non-normalizable mode
approaches $h_{\mu \nu} \rightarrow \epsilon_{\mu \nu} \frac{L^2}{R^2 v}$,
which is the four-dimensional warp factor, times a polarization tensor.

Since the \nfour\ theory is conformal in the infrared, its spectrum of
normalizable modes is continuous: any timelike $q^2$ is allowed.  We
should not think of the normalizable modes as hadrons, but simply as a
spectral decomposition of a conformal field.

\subsubsection{The discrete spectrum of hadrons of a confining model}
\label{Sec:finiteLambda}

However, it will often be useful to consider instead a
confining theory, in which the conformal invariance of the \nfour\ sector
is broken in the deep infrared.  For instance, we can imagine the
\nfour\ sector is deformed into a theory similar to \nonestar\
\cite{nonestar}, where the \none\ chiral superfields $\Phi_1,\Phi_2,
\Phi_3$ are given a small mass.  This causes confinement to set in at
a low scale $\Lambda$.  In all known models of this type, the $AdS$
radial coordinate is effectively cut off at $r=r_{min}=\Lambda R^2$.
Our results will be insensitive to the details of this cutoff, as will
become clear, so a crude model will suffice to capture the essence of
the physics. In any such model, the flavorless \nfour\ sector of the
theory will have a {\it discrete} spectrum of spin-$s$ ($s\leq 2$)
hadrons, with masses of order $n\pi\Lambda$ ($0<n\in{\bf Z}$), and with
mode functions given at large $r$ by the normalizable modes of the
corresponding bulk fields.

To be definite, we will model confinement through the boundary
condition that the ten-dimensional wave function of each hadron
satisfies the Neumann condition at $r=r_{min}$, that is, at
$v_{max}=m_h^2/\Lambda^2$.  In this simple model, the wave functions
for the hadrons created by the $SO(4)$ current are precisely those of
\eref{TheModesAn} for $v<v_{max}$, with a quantization condition that
only $q^\mu$ satisfying $(\sqrt{v_{max}} J_1(\m \sqrt{v_{max}}))'=0$, where
again $\bar m = \sqrt{|q^2|}/m_h$, are allowed.  Thus we have a
countably infinite series of modes, with mass $m_n=\m \sqrt{v_{max}} =
\zeta_{0;n}$, $n\geq 1$, where $\zeta_{0;n}$ is the $n^{th}$ zero of
$J_0$. The normalization is obtained from \cite{hardscat}
\bel{Ahadnorm} \int_0^{v_{max}} dv d^3\Omega\,
e^{2A}\sqrt{g_{\bot}}g^{MN}A_{M}A_{N}=1 , \ee
where
\[
d s^2 =e^{2A}\eta_\mn dx^\mu dx^\nu + ds_{\bot}^2 .
\]
\\
Thus we have a spectrum of states with normalized wave functions
\bel{Ahadwave} \hat A_{\mu}(m_n)= \epsilon_{\mu} \frac{\sqrt{2}
\sqrt{v/v_{max}} J_1 ( \zeta_{0;n} \sqrt{v/v_{max}})}{R^3 J_1 (
\zeta_{0;n} )}
 = \epsilon_{\mu} R^{-3}{\Lambda\over m_h}\frac{\sqrt{2v}
J_1 ( \m_n \sqrt{v})}{J_1 ( \zeta_{0;n} )} \ , \ee
and masses
\bel{Ahadmass} \bar m_n \equiv {m_n\over m_h} =
{\zeta_{0;n}}{\Lambda\over m_h}\underset{\quad n\gg
1}{\longrightarrow} \left(n-\frac14\right)\pi {\Lambda\over m_h}
\ .
\ee
We will need the nonnormalizable modes satisfying the same boundary
condition, which are (for spacelike $q^2=\q^2 m_h^2$)
\be\label{eq:so4non} \hat A_{\mu}(\q) = \epsilon_{\mu} \q \sqrt{v}
\left\{ K_1 ( \q \sqrt{v} ) + I_1 ( \q \sqrt{v} ) \frac{K_0 (\q m_h /
\Lambda )}{I_0 ( \q m_h / \Lambda )} \right\}.  \ee

Similarly, the spin-two hadrons created by
$T^{\mu\nu}$ have wave functions given by
\bel{ghadwave}
 h_{\mu\nu}(m_n)
=\epsilon_{\mu \nu} \frac{ m_h \sqrt{2/v_{max}} J_2 ( \zeta_{1;n}
\sqrt{v/v_{max}})}{R^2 J_2 ( \zeta_{1;n} )} = \epsilon_{\mu \nu}
\frac{\Lambda}{R^{2}} \frac{\sqrt{2} J_{2}(\m_{n} \sqrt{v})}
{J_{2}(\zeta_{1;n}) } \ee
with masses
\bel{ghadmass} \bar m_n \equiv {m_n\over m_h} =
{\zeta_{1;n}}{\Lambda\over m_h} \underset{\quad n\gg
1}{\longrightarrow} \left( n+\frac14 \right)\pi {\Lambda\over m_h}
\ .
\ee
where $\zeta_{1;n}$ are the zeroes of $J_1$. The normalization
constant is obtained in a similar way as in Eq. (\ref{Ahadnorm}):
\be
\int^{v_{max}}_0 dv d^3 \Omega\, e^{2A}\sqrt{g_\bot}g^{MN} g^{PQ}
h_{MP} h_{NQ} = 1 \ .
\ee
The nonnormalizable modes are now
\be\label{ghadnon}
h_{\mu\nu}(\q) = \epsilon_{\mu\nu} \frac{L^2}{R^2 v} \frac{\q^{2}
v}{2}  \left\{ K_2 ( \q \sqrt{v} ) + I_2 ( \q \sqrt{v} ) \frac{K_1
(\q m_h / \Lambda )}{I_1 ( \q m_h / \Lambda )} \right\}\ .  \ee

\section{Methodology}

\subsection{Definitions and Notation}

We will concern ourselves mainly with
the form factors from the
hadronic matrix elements of the $U(N_f)$ flavor current $J^\mu_f$,
the $SO(4)\approx
SU(2)_\Phi\times SU(2)_R$ current $J^\mu_4$, and the energy-momentum
tensor
$T^{\mu\nu}$.  We will also consider matrix elements of the
$SO(4)$-singlet spin-zero dimension-two operator
${\cal S} \propto \Phi_1^2+\Phi_2^2-2\Phi_3^2$, which is part of the
conformal \nfour\ sector.

We will compute a large number of form factors, so it is important that
our notation be clear.  When the maximal information needs to be displayed, we
will use the following notation for a form factor.  Suppose we have
a spin-zero (type $I$) or a spin-one (type $II$) initial state with $\ell, n_1$
quantum numbers.  In our calculations the final state will always
be of the same type and share the same $\ell$, though $n_2\neq n_1$
in general.  The operator whose matrix element we are computing
will be referred to by an index $S=0,1,2$ for the spin-0,1,2 operators
${\cal S}$, $J^\mu_{4}$, and $T^{\mu\nu}$, and by an index $f$ for the
flavor current $J^\mu_f$.
The maximal notation will therefore be
\bel{FnotationB}
F^{(S)\ell}_{n_1,n_2}
\ee
In many cases one or more indices will be clear from context and will
be omitted.  Finally, in computations involving spin-one initial and
final states, there can be more than one form factor.  We will label
these with an obvious subscript; in this case, the indices $S, \ell,
n_1, n_2$ will never be needed, and will be left implicit.

For scalar hadrons, the
matrix elements of the spin-one $SO(4)$ current can be written
\begin{equation}\label{eq:megeneral}
 \bra{n_2,\ell, p'} J_4^\mu ( 0 ) \ket{n_1,\ell, p} = (\eta^{\mu\nu}-q^\mu
q^\nu/q^2) (p+p')_\nu
F^{(1)\ell}_{n_1,n_2}(q^2) \ .
\ee
A similar expression holds for the flavor current.

Spin-one hadrons probed by a spin-one current have (in a
Lorentz-invariant parity-conserving theory) three form factors, one
each for electric monopole, magnetic dipole, and electric quadrupole
couplings to the spin-one current. We label these with subscripts
 $e,m,q$.  In
general the current matrix element $\bra{n_2,\ell ,p', \zeta'} J_4^{\mu}
\ket{n_1,\ell,p,\zeta }$, of spin-one mesons takes the form:
\begin{eqnarray}\label{vectorformfactors}
&& i \left\{ \right. [ ( \zeta' \cdot \zeta ) ( p' + p )_{\nu} - (
p' \cdot \zeta ) {\zeta'}_{\nu} -
( p \cdot \zeta' )\zeta_{\nu} ](\eta^{\mu\nu}-q^\mu q^\nu/q^2) F_e ( \q^2 ) + \nonumber \\
&&  + [(q \cdot \zeta' )\zeta^{\mu}-(q \cdot \zeta ) {\zeta'}^{\mu} ]F_m(\q^2)+  \\
&&  + \frac{1}{m^2}[(p \cdot \zeta' )(p' \cdot
\zeta )( p' + p )_{\nu} ](\eta^{\mu\nu}-q^\mu q^\nu/q^2) F_q(\q^2) \left. \right\} \times \nonumber \\
&& \times ( 2 \pi )^4 \delta^4 ( \sum_i p_i ) \nonumber ,
\end{eqnarray}
where $\epsilon$, $\zeta$, and $\zeta'$ are the polarization
vectors of the current, in-state, and out-state respectively, and
$F_e$, $F_m$, and $F_q$ are electric, magnetic, and quadrupole
form factors, with all other indices suppressed.
For reasons to be discussed later, in the supergravity limit
$F_e=F_m$ and $F_q=0$,
so our notation will usually be to drop the subscript except when
necessary, writing $F^{(1)\ell}_{n_1,n_2}$ for the $SO(4)$ electric
form factor and replacing $1$ with $f$ for the flavor current.

The formulas for the spin-two current are similar, although
more complicated.  Again we will find there is only one non-vanishing
form factor both for spin-zero and spin-one hadrons, which we will
label $F^{(2)\ell}_{n_1,n_2}$ when necessary.

\subsection{Determining the shape of the hadrons}

Initially we will compute the form factors in momentum space, where
they are Lorentz-invariant functions of $q^2$.  To obtain information
that is easier to interpret intuitively, we would like to reexpress
the form factors in position space.
There are ambiguities in
how this is to be done, and problems which might arise at timelike
$q^2$ where there are poles.

A four-dimensional Fourier transform of the form factor has the
feature that it is Lorentz invariant.  However this function does not
have a well-known physical interpretation.
Moreover one must consider large timelike $q^2$ where certain
difficulties with the supergravity approximation will arise.

A three-dimensional Fourier transform of the form factor at spacelike
$q^2$ is useful for nonrelativistic systems.  For a two-body
nonrelativistic bound state, this quantity is the square of the wave
function, and a similarly simple interpretation applies for many body
system: for a spin-one current, it gives the three-dimensional
distribution of the corresponding charge.  But our bound states are
highly relativistic (since their binding energy is so large), and
(for small $\ell,n$) their form factors are large  even when
$\sqrt{|q^2|}$ is of order the hadron mass.  We are therefore not
confident that this interpretation extends straightforwardly to our
case.

By contrast, the two-dimensional Fourier transform of $F(q^2)$ (for
spacelike $q^2$) does have an interpretation which is applicable for
relativistic systems.  For a hadron moving with extremely high
momentum in the $z$ direction, and choosing $q^\mu = (0,\vec{q}_\perp,
0)$ in the $x-y$ plane, we define
$$
\tilde F(x_\perp^2)
={1\over 2\pi}\int\ d^2 q_\perp\ e^{iq_\perp\cdot x_\perp} F(q^2)
$$
This function is the hadron's two-dimensional ``transverse charge
distribution'' $\rho_\perp(\vec x_\perp)$ in the $x-y$ plane, times
$2\pi$ in our conventions.  The usefulness of this interpretation
stems from its connection with generalized parton distributions
\cite{muller,ji,radyushkin}, as shown with considerable care and rigor
in \cite{burkhart}.  This applies in our case even though the (s)quarks
have large masses, as long as the hadron is ultrarelativistic compared
to the quark mass scale.\footnote{Since the
form factors are functions of $q^2$ only, one can convert from any one
of these transforms to any other, at least when restricting
to spacelike $q^2$.
Our choice of the two-dimensional transform is therefore somewhat
arbitrary; however it gives unambiguous and interpretable
information about the structure of the hadrons.}

In some cases we will find elegant closed-form
expressions for $\tilde F(x_\perp^2)$.  In others
we can still compute many of their general properties: the large- and
small-$q^2$ behavior of $F(q^2)$, the large- and small-$x_\perp^2$
behavior of its Fourier transform, and some characteristic measures of
hadron shape and size.

One classic measure of the size of a hadron is given by the second
moment $\vev{r^2}$ of the transverse charge distribution.  For any
given form-factor, we can compute this moment using
\bel{chargeradii}
\vev{r^2} \equiv 4{\partial\over \partial q^2}
F(q^2)\Big|_{q^2\to 0}
\ee
(where the 4 replaces the often-used 6 because we are measuring a {\it
two-dimensional} charge distribution.)  However, this measure is not
unique.  Another measure is $\vev{r}$ itself, which cannot so easily
be obtained from $F(q^2)$; it is best extracted directly  from the
Fourier transform $\tilde F(x_\perp^2)$.

\subsection{Calculational techniques}

The required calculational techniques are well-established and
straightforward; see for example \cite{DIS}.  Each hadron we will
consider is a mode of a particular five-dimensional field, which
itself is a mode of an eight-dimensional field on the D7 brane or of a
ten-dimensional field in the bulk.  The hadrons containing a (s)quark
and anti(s)quark will be of the so-called type $I$ or type $II$ class
described in the previous section, both of which descend from the
gauge bosons on the D7-brane.  To compute the matrix element of a
current, we need to examine the five-dimensional field whose boundary
value couples to that current.  For $J_f$, this is the mode of the
gauge boson on the D7-brane which has its index in spacetime and is
constant on the $S^3$.  (Acting on the vacuum it creates the type $II$
hadrons \Eref{TypeII} with $\ell=0$.)  For $J_4$ the gauge boson in
question is the dimensional reduction of a ten-dimensional
supergravity mode; as in \cite{DIS} it is associated with a Killing
vector on the $S^3\subset S^5$.  For $T^{\mu\nu}$ we need the
five-dimensional massless graviton.  To compute the matrix element
$\bra{out}{\cal O}\ket{in}$ requires knowlege of the trilinear
interaction between the three five-dimensional modes corresponding to
the initial hadron, the final hadron, and the operator ${\cal O}$.
This interaction can be derived from the Born-Infeld action on the D7
brane. In the supergravity limit (large $\lambda$), all of the
interactions that we will require are obtained from the single term
\bel{BIFF}
\frac1{g_8^2} \int d^8x \sqrt{-g} g^{MP}g^{NQ} e^{-D} F_{MN}F_{PQ}
\ ,
\ee
where $D$, the dilaton, will not play a role below and will be dropped
from future equations.
Here $M,N,P,Q$ are curved 8-dimensional indices, the metric
is induced from the ten-dimensional metric, and $g_8$ is the eight
dimensional Yang-Mills coupling, $(2\pi)^{5/2} \sqrt{g_s} \alpha'$.
The matrix element is then given by plugging into the appropriate
trilinear vertex a nonnormalizable mode of the field corresponding to
the current, and the wave functions of the incoming and outgoing
hadrons.
Performing the integral over the eight dimensions, we obtain
the answer required. In general, the integral over the $S^3$
and the integral over Minkowski space will be elementary.
The important integral will be that over the $AdS$ radial dimension,
and will take the form (for spin-zero incoming and outgoing hadrons)
\bel{intform}
{\rm coefficient}\ \times \int {dv\over v^2}\ \phi_{non}(q,v)\ \phi^*_{out}(v)
\ \phi_{in}(v)
\ee
where we remind the reader that $v=(L/r)^2$. For spin-one hadrons the
integrand is different only in small details.

\section{Summary of our main results}

In this section we summarize our results, and in the following section
we present the detailed computations.

\subsection{Form Factors}

\subsubsection{General results for all theories}

We begin with some observations which apparently follow from
conformal invariance and large-$N$ alone.  We have not derived them
from general arguments, but it should be possible, and would be interesting,
to do so.  In particular, it is not yet known whether the following properties
are true only in the supergravity limit and thus apply only at
large 't Hooft coupling.

The results below
apply to any hadrons with the property that (1) their
mass scale is large compared to any other scale which breaks conformal
symmetry, (2) the mass scale which sets their masses breaks conformal
symmetry only at order $1/N$.
Such hadrons will have wave functions
which solve an equation in the background of a conformal theory.  If
the conformal sector has conserved currents in addition to the
energy-momentum tensor, then we find the leading form factors of the
 energy-momentum tensor are related to those of the currents,
which are in turn related to those of operators of spin
zero and dimension two.  In our case, for scalar hadrons,
the form factors for ${\cal S}, J^\mu_4, T^{\mu\nu}$ satisfy
\bel{Fg02}
\FF{S}{}{}(\q^2) =
{(-2)^S\over S!}(\q^2)^S{\partial^S\over \partial (\q^2)^S}
\FF{0}{}{}(\q^2)
\ee
for $S=0,1,2$.
Taking a two-dimensional
 Fourier transform of these form factors
gives functions of $\xp^2$
(where $\vec\xp\equiv m_h \vec x_\perp$,)
satisfying
\bel{Fg01} \tFF{S}{}{}(\xp^2) =
{2^S\over S!}{\partial^S\over{\partial (\xp^2)^S}} \left[(\xp^2)^S
\tFF{0}{}{}(\xp^2)\right] \ .
\ee
These relations follow only from properties of the modes corresponding
to the operators in the conformal sector,
so they also apply when the incoming and outgoing hadrons
have spin one.  Note that
$$
\FF{1}{\ell}{n_1,n_2}(0) = \delta_{n_1,n_2} =
{1\over 2\pi} \int\ d^2 \xp \ \tFF{1}{}{}(\xp^2) =
{1\over 2\pi} \left[\xp^2\tFF{0}{}{}(\xp^2)\right]_{\xp=0}^{\xp=\infty} \ .
$$
Since $\tFF{0}{}{}$ is not badly
divergent\footnote{$\FF{0}{\ell}{}(q^2)\to q^{-2(\Delta_0-1)}$ at
large $q^2$, which implies $\tFF{0}{}{}$ is logarithmically divergent for
$\Delta_0=2$ and finite for larger $\Delta_0$.} at $\xp\to 0$, this
shows it must fall, at large $\xp$, as $1/\xp^2$ for $n_1=n_2$, and faster for
$n_1\neq n_2$.  One can see then that $\tFF{S}{}{}$
falls as $(\xp)^{-2(1+S)}$, or faster; the bound is saturated for
$n_1=n_2$, and also sometimes holds for $n_1\neq n_2$, $S>0$.

In contrast, the flavor current does not couple to the conformal
sector, and is sensitive to the masses of the quarks.  Its form
factors fall off
exponentially at large $\xp$.  Conformal invariance at large $q^2$
requires $\FF{f}{\ell}{n,n'} \propto\FF{1}{\ell}{n,n'} \propto
1/q^{2\ell}$; but supergravity imposes a stronger condition, namely
that the two form factors are actually {\it equal} at large $q$ (when
they are both normalized to $1$ at $q^2=0$.)

In general, spin-one hadrons can have three form factors under
spin-one currents: electric monopole, magnetic dipole, and electric
quadrupole.  We find (similarly to \cite{SonSteph}) that the anomalous
magnetic dipole ($F_m-F_e$) and the quadrupole form factor ($F_q$) are
zero in the large-$\lambda$ limit.  The nonvanishing form factor
$F_e=F_m$ has the same large-$\xp$ behavior as in the spin-zero case,
though its large $q^2$ behavior is $1/q^{2\ell+4}$.  This is true for
both $SO(4)$ and flavor currents, and follows from properties of the
couplings between hadrons and currents in the supergravity limit.
This will be discussed elsewhere \cite{sss}.

\subsubsection{General results for this theory}

In our particular theory, considerable simplification is obtained
from the fact that the wave functions in the bulk are given by
powers of $v$ and $(1-v)$ times Jacobi polynomials.
The form factors are easily analyzed in position space using
the two-dimensional Fourier transform, where they are given by
integrals of the form \eref{intform}.

In general, we find
\bel{generalform}
\tFF{S}{\ell}{n_1,n_2}(\xp^2)\propto
\Pi_1(\xp^2) \log\left(\frac{1+\xp^2}{\xp^2}\right) + \Pi_2(\xp^2)\ ,
\ee
where $\Pi_1$ and $\Pi_2$ are polynomials.
The first, up to normalization, is simply given by the wave functions
themselves, or derivatives thereof:
\bel{PiOne}
\Pi_1(\xp^2) = \frac{L^2}2
\left[\frac{\partial^S}{\partial v^S} \left(
{\phi^I_{\ell n_1}\phi^{I}_{\ell n_2}\over v^{2-S}}\right)\right]_{v=-\xp^2}
\ee
is of degree $2(\ell+n)$, and, importantly, begins at order
$(\xp^2)^{\ell-1}$.  The second polynomial $\Pi_2(\xp^2)$ is of degree
$2\ell-1$ and begins at order 1.  It can be determined by noting that
at large $\xp$ all terms which grow faster than $\xp^{-2(1+S)}$ must
cancel.  We will determine $\Pi_2$ in some specific examples.

At large $\xp$, the coefficient of the potentially-leading
$\xp^{-{2(1+S)}}$ term is
\be \int_0^1 {dv\over v^{2-S}}\ \phi^I_{\ell n_1}\phi^{I}_{\ell n_2} \ .
\ee
The above integral, for $S=0$,
is the normalization integral for the
normalizable wave functions, and it thus vanishes for $n_1\neq n_2$.
For other $S$ it need only vanish for $|n_1-n_2|\geq S$,
as follows from properties of the Jacobi polynomials.

Some of these facts have natural momentum-space counterparts.  The
behavior at small $\xp$ --- in particular the absence of a logarithm
multiplying $\xp^{2j}$ for $j<\ell-1$ --- is associated with the
requirement of conformal invariance that the momentum space form
factor fall as $1/q^{2\ell}$.  Similarly, the $(\xp^2)^{-(S+1)}$ behavior of $\tFF{S}{}{}$
at large $\xp$ follows from the fact that the expansion of $F(q^2)$
near zero is as a polynomial plus $(q^2)^S\log q$.

For spin-one hadrons, \Eref{generalform} is unmodified, but
\Eref{PiOne} becomes
\be
\Pi_{1}= \frac{R^{2}}{2} \left[\frac{\partial^S}{\partial v^S} \left(
{\frac{(1-v)\phi^{II}_{\ell n_{1}}
\phi^{II}_{\ell n_{2}}}{v^{1-S}}}\right)\right]_{v=-\xp^2}
\ee
The $\Pi_1$ polynomial now starts at order $(\xp^2)^{\ell+1}$
to account for the corresponding change in the large $q^2$ behavior,
which is now $1/q^{2\ell+4}$.

It is useful to evaluate the $SO(4)$ form factors at $\xp=0$:
\bel{tFatzero}\tFF{1}{\ell}{n_1,n_1}(0)=
\frac{2(2n_1+2\ell+2\pm1)}{\ell\pm1} \ ,
\ee
where the upper (lower)
sign applies to hadrons of spin one (zero).

The flavor form factors are less amenable to such a description
due to their mathematical complexity.  We do not have general
results beyond those of the previous section, except that all such
form factors can be written as a sum over a finite
number of spin-one $\ell=0$ hadron
poles:
\begin{equation}\label{eq:flavorffgeneral}
F^{(f)\ell}_{n_1,n_2}(\q^2 ) = \sum_{n=0}^{n_1+n_2+2\ell\mp1}
\frac{c^\ell_{n,n_1,n_2}}{\q^2+\m_n^2} \ ,
\end{equation}
where $\m_n^2=4(n+1)(n+2)$.  The $-$~($+$) sign in the upper
limit of the summation applies for spin-zero (spin-one) external
hadron states.  In position space this form factor can be
written as a corresponding sum of $K_0(m_n\xp)$ Bessel functions.

\subsubsection{Ground state form factors for each $\ell$}

Since $P_0^{(\alpha,\beta)}=1$, the ground states have
form factors proportional to
\bel{gsint}
\int_0^1 {dv\over v^2} \phi_{non}(q,v)
v^{(\elll  + 1) } ( 1 - v )^{( \elll  + 1 ) }
\ee
This implies
$$
\tFF{0}{\ell}{00}(\xp^2)=\Pi_1(\xp^2)
\log\left(\frac{\xp^2}{1+\xp^2}\right) + \Pi_2(\xp^2)\ ,
$$ \be \Pi_1=(-1)^{\ell-1}[\hat C^I_{\ell 0}]^2\ (\xp^2)^{\elll - 1 }
( 1 +\xp^2 )^{ \elll + 1 } \ee and $\Pi_2$ is proportional to the
polynomial $\mathcal{P}_{+-}$ defined in \Eref{eq:Ppmpm}. From this we
can obtain $\tFF{S}{\ell}{00}$, $S=1,2$, as described above.

For spin-one, the electric form factor has a similar form, with
\be \Pi_1=(-1)^{\ell-1}[\hat C^{II}_{\ell 0}]^2\ (\xp^2)^{\elll  +
1 } ( 1 +\xp^2 )^{ \elll  + 1  } \ee and $\Pi_2$ is proportional
to the polynomial $\mathcal{P}_{++}$ defined in \Eref{eq:Ppmpm}.

For the flavor case, $c^\ell_{n,0,0} $ in \Eref{eq:flavorffgeneral} is given in
\Eref{cspinzero} for spin zero hadrons and in \Eref{cspinone} for
spin one hadrons.  The small-$\xp$ behavior of $\tFF{f}{}{}$ is similar to
that of $\tFF{1}{}{}$.

\subsubsection{Diagonal form factors at large $\ell$}

Remarkably, although these formulas become complicated at
large $\ell$, an alternative and surprisingly simple formula
can be used instead.  We find that for both spin zero and spin one
hadrons,
\be
\tFF{S}{\ell}{00}(\xp) \approx {1\over \left(\half+\xp^2\right)^S}
\ee
for $\ell\gg 1$.
This misses the $(\xp^2)^{\ell-1}\log  \xp$ term, but this
is a negligible error at large $\ell$.
In momentum space
\be
\FF{S}{\ell}{00}(\q^2) \approx {1\over S!}(\q/\sqrt 2)^SK_S(\q/\sqrt 2)\ .
\ee

For the flavor case, the large $\ell$ limit of $c^{\ell}_{n,0,0}$ is
given in Eqs.~\eref{cspinzero-large-l} and \eref{cspinone-large-l} for
spin-zero and spin-one hadrons respectively.

\subsubsection{Diagonal form factors at large $n$}

At large $n$, a formula can be obtained which is valid for
sufficiently large $\xp$. \be \tFF{S}{\ell}{nn}(\xp) =
{1\over2^S}\binom{2S}{S} {1\over \xp(1+\xp^2)^{(2S+1)/2}} \ \
(n\gg\ell) \ee This formula applies for both spin-zero and
spin-one hadrons.  While inaccurate for very small $\xp$, these
formulas are normalizable for $S\geq 1$, and we find empirically
that this formula gives reliable answers for moments such as
$\vev{\xp}$ and $\vev{\xp^2}$.  The apparent divergence at $\xp=0$
is not present for finite $n$ (except for Type $I$, $\ell=1$) as
is clear from \Eref{tFatzero}.

For the flavor case, $c^{\ell}_{n,\infty,\infty}$ for both spin
zero and spin one hadrons is given in \Eref{cspinzero-infty}.

\subsubsection{Results for some off-diagonal matrix elements}

The conformal invariance of the \nfour\ sector
and the derivative
relations between the Jacobi polynomials imply
some additional relations
for off-diagonal matrix elements between the ground state
and an excited state at a given $\ell$.
\[ \tilde{F}^{( 0 ) \ell }_{0, n_2} ( \xp ) =
   \frac{\hat{C}^I_{\ell n_2} \hat{C}^I_{\ell 0}}{n_2 ! ( \hat{C}^I_{\ell +
   n_2, 0} )^2} \left( \frac{d}{d \xp^2} \right)^{n_2} \tilde{F}^{( 0 ) \ell +
   n_2}_{0, 0} ( \xp ) . \]
 From this expression the form factors for $S=1,2$ can be obtained
from our general result \eref{Fg01}. The same result holds for spin-one
hadrons with $\hat C^I$ replaced with $\hat C^{II}$.

\subsection{The transverse sizes of the hadrons}

Armed with this information we can compute certain moments, in
particular $\vev{r} = m_h^{-1}\vev{\xp}$ and
$\vev{r^2}=m_h^{-2}\vev{\xp^2}$, with respect to the flavor, $SO(4)$
and energy-momentum distributions of these hadrons; we use subscripts
$f$, $1$, $2$ for moments of the corresponding form factors $\FF{f}{}{}$,
$\FF{1}{}{}$, $\FF{2}{}{}$.  Where these moments are infinite we
regulate them using finite $\Lambda$ (see
Sec.~\ref{Sec:finiteLambda}); where they are finite we set $\Lambda\to
0$.

\subsubsection{The ground states $n=0$ at general $\ell$}

For every $\ell$, we can compute properties of the
ground state ($n=0$).  For the spin-zero states, we find

\[
\langle \rfour \rangle = 2\vev{\rgrav} = \frac{\pi}{2m_h} \cdot
\frac{\Gamma( \ell + \frac{1}{2} ) \Gamma( 2 \ell + 2 ) }{\Gamma(
\ell ) \Gamma( 2 \ell + \frac{5}{2} )}  \ ;
\]

\[
{\left\langle \rfour^2 \right\rangle} = \frac{1}{m_h^2} \cdot
\frac{\ell}{\ell + 1} \left\{ \log \left( \frac{m_h}{\Lambda}
\right)
\right\}\ ;
\]

\[
{\langle \rgrav^2 \rangle} = \frac{1}{8 m_h^2} \cdot
\frac{\ell}{\ell + 1} \ ;
\]

\[
\frac{0.496\dots}{m_h} \approx \langle \rflav \rangle_{\ell=1}
\leq \langle \rflav \rangle < \langle \rflav \rangle_{\ell=\infty}
\approx \frac{0.697\dots}{m_h} \ ;
\]

\[
{\langle \rflav^2 \rangle} = \frac{1}{m_{h}^2} \frac{\ell}{\ell+1}
\left[H_{2\ell+1}-H_{\ell} \right],
\]
 where $H_{\ell} = 1+1/2+1/3+\cdots+1/{\ell}$
is the $\ell$-th harmonic number.
\\

For the spin-one states, the formulas are similar:

\[
\left\langle \rfour \right\rangle = 2\vev{\rgrav}= \frac{\pi}{2
m_{h}} \frac{\Gamma(\ell+\frac52) \Gamma(2\ell+4)}{\Gamma(\ell+2)
\Gamma(2\ell+\frac92)}  \ ;
\]

\[
{\left\langle \rfour^2 \right\rangle} =
\frac{1}{m_h^2} \left\{ \log \left( \frac{m_h}{\Lambda} \right)
\right\}  \ ;
\]

\[
    \left\langle \rgrav^{2} \right\rangle =    \frac{1}{8m_{h}^{2}} \ ;
\]

\[
\frac{0.672\dots}{m_h} \approx \langle \rflav \rangle_{\ell=0}
\leq \langle \rflav \rangle < \langle \rflav \rangle_{\ell=\infty}
\approx \frac{0.697\dots}{m_h} \ ;
\]

\[
\langle \rflav^2 \rangle = \frac{1}{m_{h}^2} \frac{\ell+2}{\ell+1}
\left[H_{2\ell+3}-H_{\ell+2}\right] \ . \]

\subsubsection{The ground states at large $\ell$}

For $n=0$, $\ell \to \infty$, for both spin-zero and spin-one hadrons,

\[ \langle \rfour \rangle = 2\vev{\rgrav}
\ \longrightarrow \  \frac{\pi}{2\sqrt{2} m_{h}} \ ;
\]

\[
{\left\langle \rfour^2 \right\rangle}{\ \longrightarrow \ }
 \frac{1}{m_h^2} \left\{
\log \left(\frac{m_h}{\Lambda} \right)
\right\} \ ;
\]

\[
\vev{\rgrav^2}
{\ \longrightarrow \ } \frac{1}{8 m_h^2}\ ;
\]

\[
 \langle \rflav \rangle \ \longrightarrow\
  {0.697\dots\over m_h}\ ;
\]

\[
\langle \rflav^2 \rangle \ \longrightarrow \ {\log2\over m_h^2} \
.
\]

\subsubsection{The small-$\ell$ states at large $n$}

For both spin-zero and spin-one hadrons, the limit $n\to \infty$
at small and fixed $\ell$ gives

\[
{\left\langle \rfour \right\rangle} = 2\vev{\rgrav} {\
\longrightarrow \ } \frac{1}{m_{h}} \ ;
\]

\[
\left\langle \rfour^2 \right\rangle {\ \longrightarrow \ }
\frac{1}{m_h^2} \left \{ \log\left (
\frac{m_h}{\Lambda} \right )
\right\}\ ;
\]

\[
{\langle \rgrav^2 \rangle}  {\ \longrightarrow \ } \frac{1}{8
m_{h}^2} \ ;
\]

\[
\vev{r_f} {\ \longrightarrow \ } \frac{0.61\dots}{m_h} \ ;
\]

\[
\vev{\rflav^2} {\ \longrightarrow \ } {2(1-\log2)\over m_h^2} \ .
\]

\section{Computations}

In this section we derive the results outlined in the previous section.

\subsection{Spin zero hadrons}
We now proceed to study the spin-zero (type $I$) hadrons created by
the operators $\tilde Q\Phi^{\ell-1} Q$ (and other members of the
same $SO(4)$ multiplet.)

\subsubsection{Flavor current}

 In order to study the (s)quark-anti(s)quark hadrons
 using the matrix elements of the flavor current, we need
to consider a situation with $N_f>1$.  The hadrons in question
transform in the adjoint of $U(N_f)$, so for $N_f=1$ they are
neutral under the flavor current.  We will consider $N_f$
hypermultiplets of equal mass, which leaves the $U(N_f)$ unbroken.
The modes described in \cite{Myers}, which only depend on the
quadratic terms in the D7-brane gauge fields, are unchanged for
$N_f>1$, except for transforming under a nontrivial representation
of flavor. However, the cubic interactions among the D7-brane
gauge bosons give the main contribution to the flavor-current
matrix elements.

For the coupling of the flavor current to two spin-zero hadrons,
the important term in the D7-brane Born-Infeld action is
\[g_8 \int d^8x \sqrt{-g}
g^{\alpha\beta}g^{\mu\nu} f^{abc} A^a_{{\mu}}A^b_{{\alpha}}
\partial_{\nu}A^c_{\beta},\] where $f^{abc}$ is the structure
constant of the group $SU(N_f)$ and we have rescaled the gauge
fields $A^M \rightarrow g_8 A^M$ to obtain this form. In this
form, the boundary ($r = \infty$) value of the non-normalizable
mode times the coupling $g_8$ is set to unity.

Substituting the non-normalizable mode for $A_{\mu}$, and
normalizable modes for the mesons, we obtain the matrix element:

\[
\bra{a;\ell,n_2;p'}{J_f^b}^{\mu}(q)\ket{c;\ell,n_1;p} =  i f^{abc}
( p + p' )_\nu (\eta^{\mu\nu}-q^\mu q^\nu/q^2) ( 2 \pi )^4
\delta^4 \left( \sum_i p_i\right) F^\ell_{n_1,n_2} (\q^2 ),
\]

\begin{equation}
F^\ell_{n_1,n_2} (\q^2 ) = \frac{L^2}{2} \int_0^1 dw (1-w)^{-2}
\phi^{II}_{non} (w;\alpha) \phi^{I}_{\ell,n_1} (w)
\phi^{I}_{\ell,n_2} (w) . \label{F}
\end{equation}

The above integral can be done by partial integration, using the
modes in \Eref{TypeI} and \Eref{flavornon}. The normalizable mode
is a polynomial in $w$ and $\int dw F(a+1,b+1;c+1;w)=
\frac{c}{ab}F(a,b;c;w)$. By integrating the non-normalizable mode
and differentiating the product of normalizable modes repeatedly,
this integral can be evaluated. The general form of the form
factor obtained in this way is
\begin{equation}
F^\ell_{n_1,n_2}(\q^2 ) = \sum_{n=0}^{n_1+n_2+2\ell-1}
\frac{c^\ell_{n,n_1,n_2}}{\q^2+\m_n^2},\label{ff}
\end{equation}
 where $\m_n^2=4(n+1)(n+2)$ is the  squared mass of the $n$-th vector
meson with $\ell=0$, and $c_{n,n_1,n_2}$ is a constant independent of $\q$.

For the general case, we can get the large $\q^2$ behavior in the
following way: Remembering that we evaluate the integral by
repeated partial integration, the $\ell$-times integration of
$\phi^{II}_{non} (w;\alpha)$ gives a term with its leading
behavior $4^{\ell} \ell! / (\q^2)^{\ell}$ for large $\q^2$ and the
function we need to differentiate repeatedly has the structure
$${L^2\over(1-w)^{2}}\phi^{I}_{\ell,n_1} (w) \phi^{I}_{\ell,n_2} (w) =
\frac{\ell(\ell+1)C^I_{\ell n_1}}{(n_1+\ell)(n_1+\ell+1)}
\frac{\ell(\ell+1)C^I_{\ell n_2} }{(n_2+\ell)(n_2+\ell+1)}
w^{\ell+1}(1-w)^{\ell-1}$$
plus corrections suppressed by additional factors of
$(1-w)$.  Therefore, the first
$\ell-1$ terms obtained by partial integrations vanish and the
leading large $\q^2$ behavior is given by
\bel{eq:Spin0LargeQ}
F^\ell_{n_1,n_2}(\q^{2}) \underset{\quad \q^2 \to \infty}{\longrightarrow}
\frac{1}{2} \frac{\ell(\ell+1)C^I_{\ell
n_1}}{(n_1+\ell)(n_1+\ell+1)} \frac{\ell(\ell+1)C^I_{\ell
n_2}}{(n_2+\ell)(n_2+\ell+1)} \frac{ 4^{\ell}
\ell!}{(\q^2)^{\ell}} \propto \frac{1}{(\q^2)^{\Delta_0 -1} },
\ee
where $\Delta_0 = \ell+1$ is the conformal dimension of the
spin-$0$ mode. This behavior agrees with the ``quark counting
rules'' in QCD \cite{brodskyfarrar,mmu}, because the latter follow
from conformal invariance and are not in fact limited to weak
coupling and the valence-parton model of hadrons.

It is possible to compute the Fourier transformation of the form
factor. Using the 2-d Fourier transformation
\[
FT\left(\frac{1}{q^2+m^2}\right) = K_0(m \xp),
\]
we get
\begin{eqnarray}
\tilde F^\ell_{n_1,n_2}(\xp) & = & m_h^2 \sum_{n=0}^{n_1+n_2+2\ell-1} c^\ell_{n,n_1,n_2}K_0(m_n \xp) \nonumber \\
& \underset{\xp\to\infty}{\longrightarrow} & m_h^2 c^\ell_{\hat{n} ,n_1,n_2}
\sqrt{\frac{\pi}{2m_{\hat{n}} \xp}} \quad e^{-m_{\hat{n}}
\xp},
\label{FTff}\\
& \underset{\xp\to 0}{\longrightarrow} &  m_h^2
 \!\!\!\! \sum_{n=0}^{n_1+n_2+2\ell-1} \!\!\!\! c^\ell_{n,n_1,n_2}
\sum_{k=0}^{\infty} \left[-\log{\left(\frac{m_n \xp}{2}\right)}
+ \psi(k+1)\right]
\frac{\left(\frac{m_n \xp}{2}\right)^{2k}}{(k!)^2},
\label{eq:Spin0FT}
\end{eqnarray}
%
 where
$\hat{n}$ is the smallest value of $n$ with nonvanishing
$c^\ell_{n ,n_1,n_2}$. If $\sum c^\ell_{n,n_1,n_2}=0$, the
logarithmic divergence for $\xp \rightarrow 0$ vanishes; as
can be seen from \Eref{ff}, this is also the condition that
the coefficient of $1/\q^2$ at large $\q^2$ vanishes.
We know this must occur for $\ell>1$.
In general, from the fact that the form factor falls as
$1/\q^{2\ell}$, we can deduce that $\sum
c^\ell_{n,n_1,n_2} (\m_n)^{2j} = 0$ for $j=0,1,\cdots, \ell -2$, and
thus the leading non-analytic term in the small-$\xp$ 
expansion is
$-[\xp^{2(\ell-1)}+ \cdots]\log{\xp}$. 

In principle, we can get all $c^\ell_{n,n_1,n_2}$ by partial
integrations, but it's difficult to get a closed form for
$c^\ell_{n,n_1,n_2}$ in general. In the case of
$n_1=n_2=0$, we can get a closed form\footnote{This form can be 
obtained by using the decomposition $c_{n,0,0}^{\ell} = f_n
g_{n,0,0}^{\ell}/m_h^2$, which is briefly discussed in the appendix.}
which is given by
\begin{eqnarray}
c^\ell_{n,0,0}
&=&  {C^{II}_{0,n}}^2 {C^{I}_{l,0}}^2 B(\ell+1,n+\ell+2){}_3F_2(-n,-n-1,\ell+1;2,-n-\ell-1;1) \nonumber\\
&=& \left\{ \begin{array}{ll}
            (-1)^{\frac{n}{2}} \phantom{0} \frac{4(2n+3)(n+1)!}{[(n/2)!]^2}
                \frac{(1+2\ell)!(\ell+n/2)!}{(\ell+1)(\ell-n/2-1)!(1+2\ell+n)!}
                & \text{$n$  even}\\ \\
            (-1)^{\frac{n-1}{2}} \phantom{0} \frac{2(n+1)(2n+3)(n+2)!}{[(n/2+1/2)!]^2}
                \frac{(1+2\ell)!(1/2+\ell+n/2)!}{(\ell+1)(\ell-n/2-1/2)!(2+2\ell+n)!}
                & \text{$n$  odd}
            \end{array}
    \right. \label{cspinzero} \\ \nonumber \\
&\rightarrow& \left\{ \begin{array}{ll}
                        (-1)^{n/2} \phantom{0} \frac{4(2n+3)(n+1)!}{2^n [(n/2)!]^2}
                            & \text{$n$ even} \\ \\
                        {\rm Order}(1/\ell) \rightarrow 0
                            & \text{$n$  odd}
                        \end{array}
                ,\quad \ell \rightarrow \infty
              \right. \label{cspinzero-large-l}
\end{eqnarray}
where $B(a,b)$ is the Beta function and we used the Stirling's
formula $z!\sim \sqrt{2\pi} z^{z+\half}e^{-z}$ in the last
equation.

In the case of $n_1=n_2 \rightarrow \infty$, $c_{n,n_1,n_2}^{\ell}$,
becoming independent of $\ell$, is given by\footnote{Here we used the
  asymptotic representation of the Jacobi polynomial,
\[
P_n^{(\alpha, \beta)} (\cos\theta) = \frac{\cos \left( [n+
(\alpha+\beta+1)/2]\theta - (\alpha /2 +1/4)\pi \right) }{
\sqrt{\pi n} \left( \sin\frac{\theta}{2} \right)^{\alpha+ \half}
\left(\cos\frac{\theta}{2} \right)^{\beta + \half} } + O(n^{-
\frac{3}{2}}), \qquad 0<\theta<\pi
\]}
\begin{eqnarray}\label{cspinzero-infty}
c^\ell_{n,\infty,\infty}
&=&  {C^{II}_{0,n}}^2 \frac{(2n)!}{2^{2n}n!(n+1)!} \phantom{0} {}_3F_2(-n,-n-1,\frac{3}{2};2,\half -n;1) \nonumber\\
&=& \left\{ \begin{array}{ll}
             + \frac{(2n+3)[(n+1)!]^2}{2^{2n-2}[(n/2)!]^4}
                & \text{$n$  even}\\ \\
             - \frac{(n+1)(2n+3)(n+1)!(n+2)!}{2^{2n}[(n/2+1/2)!]^4}
                & \text{$n$  odd}
            \end{array}
    \right. \\ \nonumber \\
&\rightarrow& (-1)^n \frac{16}{\pi}n^2  ,\quad n \rightarrow
\infty \nonumber
\end{eqnarray}

We can use this information to get some measures
of the size of the scalar meson.  First, we can get an
exact answer for
$(\ell,n)=(\infty,0)$,
\begin{eqnarray*}
m_h^2 \langle \rflav^2 \rangle_{\ell\to\infty}
&=& 4 \sum_{n=0}^{\infty}\frac{c^{\ell = \infty}_{n,0,0}}{(\bar{m}_n)^4}\\
&=& \sum_{k=0} (-1)^k \frac{(2k)!}{2^{2k}(k!)^2(2k+1)}-\sum_{k=0}
(-1)^k \frac{(2k+1)!}{2^{2(k+1)}[(k+1)!]^2}\\
&=& \sinh^{-1}(1) - [\log(1+\sqrt{2})-\log2]
\ =\  \log2
\end{eqnarray*}
Next, $\langle \rflav \rangle_{\ell\to\infty}$ can be estimated
well enough numerically by considering the first few terms in the
summation because $c^{\ell = \infty}_{n,0,0}$ has alternating sign
and the magnitude of ${c^{\ell = \infty}_{n,0,0}}/{\m_n^3}$
decreases for large $n$:  for large even $n$,
\[
\frac{c^{\ell =
\infty}_{n,0,0}}{\m_n^3} \sim (-1)^{n/2} \frac{2}{\sqrt{2\pi}}
n^{-3/2} \ .
\]
This gives
\[
m_h \langle \rflav \rangle_{\ell\to\infty} =
\frac{\pi}{2}\sum_{n=0}^{\infty}\frac{c^{\ell =
\infty}_{n,0,0}}{(\bar{m}_n)^3} \approx 0.697 \ .
\]
For general $\ell$,
\[
m_h \langle \rflav \rangle_{\ell} =
\frac{\pi}{2}\sum_{n=0}^{2\ell-1}
\frac{c_{n,0,0}^{\ell}}{(\m_n)^3} \qquad ,
\]
which increases with $\ell$ but is bounded by $\langle \rflav
\rangle_{\ell\to\infty}$.
We can also estimate $\langle \rflav \rangle_{n\to\infty}$ by
using $c_{n,\infty,\infty}^{\ell}$ for $n_1=n_2=\infty$:

\be m_h \langle \rflav \rangle_{n\to\infty} \approx 0.61 \ . \ee

More exact results on $\langle \rflav^2 \rangle_{\ell}$ can be
obtained by differentiating
\Eref{F} with respect to $q^2$ first and then integrating. For the
scalar meson with $(\ell,n=0)$, \bel{rfground} {\langle \rflav^2
\rangle_{\ell}} = \frac{1}{m_{h}^2}
\frac{\ell}{\ell+1}\left[\psi(2\ell+2)-\psi(\ell+1)\right] =
\frac{1}{m_{h}^2} \frac{\ell}{\ell+1} \left[H_{2\ell+1}-H_{\ell}
\right], \ee
 where
$H_{\ell} = 1+1/2+1/3+\cdots+1/{\ell}$ is the $\ell$-th harmonic
number. For example, $ \langle \rflav^2 \rangle_{\ell=1} = 5/
12m_h^2$ and the squared radius is bounded by \bel{rfllarge}
\langle \rflav^2 \rangle_{\ell\to\infty} = {\log2\over m_h^2} \
.\ee The effect of large $n$ is to make $\langle \rflav^2 \rangle$
decrease to some extent, but $\langle \rflav^2 \rangle$ is bounded
from below :

\bel{rfnlarge}\langle \rflav^2 \rangle_{n\to\infty} =
{2(1-\log2)\over m_h^2} \ . \ee
\\

\noindent\underline{{\bf Examples:}}

\

In the case of $\bra{1,0} J_f^{\mu}(q) \ket{1,0}$,
\[F_{0,0}^{\ell=1} (\q^2 ) =
\frac{6}{\q^2+\m_{0}^2}+\frac{6}{\q^2+\m_{1}^2} \sim
\frac{12}{\q^2}
\]
\begin{eqnarray*}
\tilde{F}_{0,0}^{\ell=1}(\xp)
& = & 6 m_h^2 \left[K_0(m_{0}\xp)+K_0 (m_{1} \xp) \right] \\
& \rightarrow & 6 m_h^2\sqrt{\frac{\pi}{2m_0\xp}} \quad e^{-m_0 \xp}, \qquad \xp \rightarrow \infty \\
& \rightarrow & - 12 m_h^2 \log{(m_h \xp)}, \qquad \xp \rightarrow
0
\end{eqnarray*}
The logarithmic divergence for $\xp \rightarrow 0$ stems from the
leading $1/q^2$ behavior for large $\q^2$.

\rem{ For $\bra{1,1} J_f^{\mu}(q) \ket{1,1}$,
\[F_{1,1}^{\ell=1} (\q^2 ) =
\frac{10}{\q^2+\m_{0}^2}- \frac{90/7}{\q^2+\m_{1}^2} +
\frac{160/7}{\q^2+\m_{3}^2}\sim \frac{20}{\q^2}
\]

\[
\tilde{F}_{1,1}^{\ell=1}(\xp) = m_h^2 \left[10 K_0(m_{0} \xp)
-(90/7)K_0 (m_{1}\xp)+ (160/7)K_0 (m_{3}\xp) \right]
\]

\begin{eqnarray*}
\tilde{F}_{1,1}^{\ell=1}(\xp) & \rightarrow & 10 m_h^2
\sqrt{\frac{\pi}{2m_0
\xp}} \quad e^{-m_0 \xp}, \qquad \xp \rightarrow \infty \\
& \rightarrow & -20 m_h^2 \log{(m_h \xp)}, \qquad \xp \rightarrow 0
\end{eqnarray*}

For $\bra{2,0} J_f^{\mu}(q) \ket{2,0}$,

\[
F_{0,0}^{\ell=2}(\q^2)= \frac{8}{\q^2+\m_0^2}
+\frac{40/7}{\q^2+\m_1^2}-\frac{8}{\q^2+\m_2^2}-\frac{40/7}{\q^2+\m_3^2}
\sim \frac{640}{\q^4}
\]
\[
\tilde{F}_{0,0}^{\ell=2}(\xp) = m_h^2[8 K_0(m_0 \xp)+(40/7)K_0(m_1
\xp)- 8 K_0(m_2 \xp)- (40/7) K_0(m_3 \xp)]
\]
\begin{eqnarray*}
\tilde{F}_{0,0}^{\ell=2}(\xp)& \rightarrow & 8 m_h^2
\sqrt{\frac{\pi}{2m_0 \xp}} \quad e^{-m_0 \xp}, \qquad \xp \rightarrow \infty \\
& \rightarrow & m_h^2 [ 4 \log{6} + (20/7) \log{(10/3)} ], \qquad
\xp \rightarrow 0
\end{eqnarray*}

Since $F_{n_1,n_2}^{\ell=2}(\q^2)$ falls faster than $1/\q^2$ for large $\q^2$,
$\tilde{F}^{\ell=2}_{0,0}(\xp^2)\rightarrow \mbox{constant}$ for $\xp
\rightarrow 0$.
}

More generally,
\[
\FF{f}{\ell=1}{0,n_2} (\q^2 ) =
(-1)^{n_2+1}\sqrt{\frac{12}{2n_2+3}}\left[\frac{n_2(n_2+2)}{\q^2+\m_{n_2-1}^2}
-\frac{2n_2+3}{\q^2+\m_{n_2}^2}-\frac{(n_2+1)(n_2+3)}{\q^2+\m_{n_2+1}^2}
\right].
\]In the limit of large $\q^2$, the leading behavior of
this form factor goes like $1/\q^2$.

\subsubsection{$SO(4)$ current}

The $SO(4)$ current is associated with the
Killing vectors which generate the isometry of $S^{3}$.
For simplicity we limit ourselves
to the two $U(1)$ subgroups in the maximal
torus of $SO(4)$ under which our hadrons are eigenstates,
namely the diagonal $U(1)$ subgroups
of $SU(2)_{\Phi}$ and
$SU(2)_{R}$.

In this case, the setup is almost identical to the one
in~\cite{DIS}.
We consider the interaction of the hadron modes
with the canonically normalized $SO(4)$ mode on the D7 brane,
given in section~\ref{sec:N4sector}, through the vertex
\[ \frac{\kappa}{R}\int \sqrt{g} \hat A^m v^{\alpha} g^{\beta \gamma}
\partial_m A_{\beta}
   \partial_{\alpha} A_{\gamma}
   = \frac{\kappa}{R} i \mathcal{Q} \int \sqrt{g} \hat A^m g^{\beta
   \gamma} \left\{ ( \partial_m A_{\beta} )_f^{\ast} ( A_{\gamma} )_i
   - ( A_{\beta} )^{\ast}_f ( \partial_m A_{\gamma}^{} )_i \right\} \ . \]
where $i \mathcal{Q}$ is the eigenvalue from acting with
the Killing vector $v^\alpha \partial_{\alpha}$
on the mode.

When the non-normalizable mode with canonical normalization is put
in this vertex along with the normalizable modes, we have the
matrix element
\begin{equation}
\bra{\ell,n_2;p'} J_4^\mu\ket{\ell,n_1;p} =  \mathcal{Q}
\epsilon^\mu ( p + p' )^\nu (\eta_{\mu\nu}-q_\mu q_\nu/q^2) ( 2
\pi )^4 \delta^4 \left( \sum_i p_i \right) F_{n_1,n_2}^{(1)\ell} (
\q^2 ) \ , \ee
\begin{equation}\label{eq:mespin0}
\quad \, F_{n_1,n_2}^{(1)\ell} ( \q^2 ) \, = \frac{L^2}{2}
\int^1_0 \frac{d v}{v^2}\, \q \sqrt{v} K_1 ( \q \sqrt{v} )
\phi^{I}_{\ell,n_1} ( v ) \phi^{I}_{\ell,n_2} ( v ) \ .
\end{equation}
Note that \eref{eq:mespin0} approaches the normalization integral
for the normalizable modes in the $q^2 \to 0$ limit. This
guarantees that $F^{(1)\ell}_{n_1,n_2}(0) = \delta_{n_1,n_2}$.

Since $\phi^{I}_{\ell,n} ( v )$ is a polynomial in $\sqrt{v}$, the exact
form of the integral can be obtained for arbitrary modes in
principle, but the general form is hard to obtain. However, we can
obtain some interesting results which apply to all or many of the states.
First of all, as in the flavor case, the form factor goes like
$1/(q^2)^{\Delta_0-1}$. From~\Eref{eq:mespin0},
$\q \sqrt{v} K_1 ( \q \sqrt{v} )$ falls to zero very rapidly and
the integration gets the contribution mostly from the region $0
\leq v \lesssim 1/\q^2$. It is therefore useful to convert the
integration variable to $\q^2 v$. Now, suppose $q^{2}\gg m_h^2$ is
large $(\q^2\gg 1)$.  Then the term with the lowest power of
$\q\sqrt{v}$ yields the leading contribution. The calculation only with
the term of the minimal power in $v$ is easy and it gives us
exactly~\Eref{eq:Spin0LargeQ}.\footnote{This is not a surprise.
In the large $q^{2}$ limit, the form factor is not affected by the
quark masses and is governed by conformal invariance. In other
words, the equation for the flavor mode can be written as
\[
(\q^{2} v) \frac{d^{2}}{d(\q^{2} v)^{2}} G + \frac1{4} G
= \frac{1}{\q^{2}}\frac{d}{d(\q^{2} v)}\left( (\q^{2} v)^{2}
\frac{d}{d(\q^{2} v)} G \right)
\]
Therefore, we can consistently expand $G(\q,v) = G_{0}
(\q \sqrt{v}) + G_{1}(\q\sqrt{v})/\q^{2} + \cdots$ where
$G_{0}$ is precisely the conformal mode~\eref{TheModesAnn}.}

To compute the two-dimensional Fourier transformation of the form factor is
straightforward. \be\label{eq:frmeso4general}
\tilde{F}_{n_1,n_2}^{(1)\ell} ( \xp^2) = \frac{L^2}{2} \int
\frac{d v}{v^2} \, \frac{2 v}{(v+\xp^2 )^2} \cdot \phi^{I }_{\ell,
n_1} ( v ) \phi^{I }_{\ell, n_2} ( v ) \ . \ee Again, we can
extract the important information without getting into too much
computational detail. First of all, instead of the previous
equation, we consider the following

\be \tilde{F}^{(0)} ( \xp^2 ) = \frac{L^2}{2} \int^1_0 \frac{d
v}{v^2} \,
   \frac{\phi^I_{\ell n_1} \phi^I_{\ell n_2}}{v + \xp^2} \ ,
   \label{eq:spin0F0}
\ee
which has the following relationship with the Fourier
transformation we want: \be \tilde{F}^{(1)} ( \xp^2 ) = 2
\frac{d}{d \xp^2} \xp^2 \tilde{F}^{(0)} (
   \xp^2) \ . \label{eq:F0toF1}
\ee
This is proven in the appendix, as are the more general set of relations
\eref{Fg02} and \eref{Fg01}.
Now, by partial integration, we have
\[ \tilde{F}^{( 0 )\ell}_{n_{1},n_{2}} ( \xp^2 ) =
\Pi_1 ( \xp^2 ) \log \left( 1 + \frac{1}{\xp^2}
   \right) + \Pi_2 ( \xp^2 ) \]

\[ \Pi_1 ( \xp^2 ) = \left. \frac{L^2}{2} \frac{\phi^I_{\ell n_1} \phi^I_{\ell
   n_2}}{v^2} \right|_{v = - \xp^2} \]
In the small-$\xp$ region, $\Pi_2(\xp^2)$ dominates because the
lowest order term of $\Pi_1(\xp^2)$ is $\sim (\xp^2)^{\ell-1}$,
except for the case $\ell=1$ for which there is a logarithmic
singularity as $\xp\to0$. This behavior is the same as in the
flavor form factors. Once again, this is not a coincidence since
it is related to the behavior of the form factors at large $q^2$,
which should be the same for both cases by conformal symmetry.

The large-$\xp$ behavior is obtained by approximating $(v+\xp^2)
\to \xp^2$ in \Eref{eq:frmeso4general}.  The coefficient of the
leading $1/\xp^4$ term turns out to be non-zero only for $n_1=n_2$
and $n_1=n_2 \pm 1$ because of the orthogonality and the
recurrence relations of Jacobi polynomials.

We also compute the $SO(4)$ charge radius squared $\langle \rfour^{2}
\rangle$ of the mesons  as given in \eref{chargeradii}.  In the
purely conformal case $\Lambda=0$, we find a logarithmic divergence.
This is an infrared effect, stemming from the continuous spectrum of
the conformal field theory in the \nfour\ sector.  The introduction of
a nonzero $\Lambda$ regulates the divergence.  The non-normalizable
modes for $\hat A$ are now those of \Eref{eq:so4non}, and so we obtain

\begin{equation}\label{eq:so4radker}
\left. \frac{d}{d q^2} \q \sqrt{v}\left\{ K_1 ( \q \sqrt{v} ) +
I_1 ( \q \sqrt{v})
 \frac{K_0 (\q m_h / \Lambda )}{I_0 (\q m_h / \Lambda )}
 \right\} \right|_{q^2 = 0} = 
 -\frac{v}{2 m_h^2}
 \left\{ \log \left( \frac{m_h}{\Lambda} \right) - \frac12 \log v +
 \frac14
 \right\}. \nonumber
\end{equation}
Thus $\left\langle \rfour^2 \right\rangle$ can be written in closed
form for the ground state meson $n_{1}=n_{2}=0$ for any $\ell$.
\bel{eq:rfour2gdstate} 
{\left\langle \rfour^2 \right\rangle} =
\frac{1}{m_h^2} \cdot \frac{\ell}{\ell + 1} \left\{ \log \left(
\frac{m_h}{\Lambda} \right) + \half ( H_{2 \ell + 2} - H_\ell ) +
\frac{1}{4}
\right\}.
\ee
Note that this has the earlier-noted logarithmic
divergence in the limit $\Lambda\to 0$ with $m_h$ fixed, but goes
to zero in the limit $m_h\to \infty$ with $\Lambda$ fixed.

For large $\ell$, $n_1=n_2=0$, we find
\bel{eq:so4r2largel}
{\left\langle \rfour^2 \right\rangle}
\underset{\quad l \rightarrow \infty}{\longrightarrow}
 \frac{1}{m_h^2} \left\{
\log \left(\frac{m_h}{\Lambda} \right) + \half \log 2 + \frac14 
   \right\} \ ,
\ee which shows the same divergence as before. As $n_1=n_2$
becomes large, $\ell$ fixed, \be\label{eq:so4r2largen}
\left\langle \rfour^2 \right\rangle \underset{\quad
n_{1}\to\infty} {\longrightarrow} \frac{1}{m_h^2} \left \{
\log\left ( \frac{m_h}{\Lambda} \right ) +\log 2+\frac{3}{4}
\right\} \ , \ee
which also shows the same logarithmic divergence.

However, the logarithmic divergence is
misleading.  It is merely an indication that
${\left\langle \rfour^2 \right\rangle}$ is not a good measure in
this theory. Since $\q \sqrt{v} K_{1}(\q \sqrt{v}) = 1 + \q^{2}v
\log{(\q^2 v)}/4 +\cdots$ when $\q\sqrt{v}\ll 1$, the two dimensional Fourier
transform of the second term is
\[ \int d^2 \q \, e^{i \q \cdot \xp}  \q^2 v \log{(\q^2 v)}/4 \sim
\frac{2 v}{\xp^{4}} + \text{(UV sensitive terms)} \ .
\]
This $1/\xp^4$ tail leads to a logarithmic divergence, since
\[ \langle r^2_1 \rangle \sim \frac{1}{m_h^2} \int d^2 \xp \, \ \xp^2
   \cdot \frac{1}{\xp^{4}} \sim \frac{1}{m_h^2} \log ( m_h /
   \Lambda ) \ .
\]

On the other hand, there are other measures of the size which don't suffer
from this divergence. For example, from the previous equation
it is clear that $\vev{\rfour}$ is finite.  Using
\[
 \int^{\infty}_0 d \xp \, \frac{\xp^2}{( \xp^2 + v )^2} = \frac{\pi}{4
   \sqrt{v}} \ , \]
we can show that for $n=0$ and general $\ell$,
\bel{eq:vevrfour}
\langle \rfour \rangle = \frac{\pi}{2m_h} \cdot \frac{\Gamma( \ell +
   \frac{1}{2} ) \Gamma( 2 \ell + 2 ) }{\Gamma( \ell )
\Gamma( 2 \ell + \frac{5}{2} )} \ , \ee which has a limit
\bel{eq:so4r1largel} \langle \rfour \rangle \to \pi/(2\sqrt{2}
m_{h}) \ \ (\ell \to \infty) \ .  \ee
This result is not significantly modified for the excited states:
\be
    {\left\langle \rfour \right\rangle} \underset{\quad n_{1} \to
\infty}{\longrightarrow}  \frac{1}{m_{h}} \ .\label{eq:so4rlargen2}
\ee

\

\noindent\underline{{\bf Examples:}}

\

We begin with the form factor of the matrix element $\bra{\ell,
n_1 = 0} J^{\mu} \ket{\ell, n_2}$ in the position space. As
explained before, the computation of  $\tilde{F}^{( 0 )}$ is much
easier in many cases and the higher spin form factors can be
derived from it. We easily obtain
\begin{eqnarray*}
  \tilde{F}^{( 0 ) \ell }_{0, n_2} ( \xp^2 )_{} & = & \hat{C}^I_{\ell n_2}
  \hat{C}^I_{\ell 0} \frac{( - 1 )^{n_2}}{n_2 !} \int^1_0 dv \frac{1}{v +
  \xp^2} \left( \frac{d}{dv} \right)^{n_2} ( 1 - v )^{n_2 + \ell + 1} v^{n_2 +
  \ell - 1}\\
  & = & \frac{\hat{C}^I_{\ell n_2} \hat{C}^I_{\ell 0}}{n_2 !} \int^1_0 dv
  \hspace{0.25em} ( 1 - v )^{n_2 + \ell + 1} v^{n_2 + \ell - 1} \left(
  \frac{d}{dv} \right)^{n_2} \frac{1}{v + \xp^2}\\
  & = & \frac{\hat{C}^I_{\ell n_2} \hat{C}^I_{\ell 0}}{n_2 !} \left(
  \frac{d}{d \xp^2} \right)^{n_2} \int^1_0 dv \hspace{0.25em} \frac{( 1 - v
  )^{n_2 + \ell + 1} v^{n_2 + \ell - 1}}{v + \xp^2} \ .
\end{eqnarray*}
Here we discover an identity, \be\label{eq:F00toF0n} \tilde{F}^{(
0 ) \ell }_{0, n_2} ( \xp^2 ) =
   \frac{\hat{C}^I_{\ell n_2} \hat{C}^I_{\ell 0}}{n_2 ! ( \hat{C}^I_{\ell +
   n_2, 0} )^2} \left( \frac{d}{d \xp^2} \right)^{n_2} \tilde{F}^{( 0 ) \ell +
   n_2}_{0, 0} ( \xp^2 ) \ ,
\ee
which is derived using a relation proven in the appendix.
The rest of the computation is straightfoward,
\[ \tilde{F}^{( 0 ) \ell }_{ 0, 0}(\xp^2) = ( \hat{C}^I_{\ell 0}
   )^{2} \int_0^1 dv \frac{( 1 - v )^{\ell + 1}
   v^{\ell - 1}}{v + \xp^2} = ( \hat{C}^I_{\ell 0} )^2 \left[ \hat{\Pi}_1 ( \xp^2 )
   \log \left( 1 + \frac{1}{\xp^2} \right) + \hat{\Pi}_2 ( \xp^2 ) \right], \]
\[ \hat{\Pi}_1 = ( 1 + \xp^2 )^{\ell + 1} ( - \xp^{2} )^{\ell - 1}, \quad
\hat{\Pi}_2 =\mathcal{P}_{+ -} ( \xp^2 ), \]
\begin{multline}\label{eq:Ppmpm}
\mathcal{P}_{+ \pm} ( y ) =
\pm \sum^{\ell \pm 1 - 1}_{j = 0} \binom{2 \ell
   - j \pm 1 + 1}{\ell \pm 1}^{- 1} \frac{( - y )^j}{\ell \pm 1 - j}
\\
   - ( - 1 )^{\ell \pm 1} \sum_{j = \ell \pm 1 }^{2 \ell \pm 1 }
   \binom{\ell + 1}{j - \ell \mp 1 } ( H_{\ell + 1} - H_{j - \ell \mp
   1} ) y^j .
\end{multline}
As an example, the diagonal form factor for the lowest lying state is
\[ \tilde{F}^{( 1 ) \ell = 1 }_{0, 0} ( \xp^2 ) = - 3 ( 5 + 6 \xp^2 ) +
6  \frac{d}{d \xp^2} \left[\xp^2 ( 1 + \xp^2 )^2 \right] \log ( 1 + 1 /
\xp^2 ) .  \]

Note that $F^{(0)\ell}_{0,0}$ has an interesting limit as $\ell$ goes to
infinity.  Since the minimal power of the polynomial multiplying the
log is at least $v^{\ell-1}$, the log part makes the subleading
contribution in that limit.  This is the same for the second summation
in \eref{eq:Ppmpm}.  The coefficient in the first term multiplied by
$(\hat{C}^{I}_{\ell 0})^{2}$ is simplified greatly $(\ell -
1-j)\binom{2\ell -j-1}{\ell\pm1} \longrightarrow 2^{-j-1}$. 
Hence, the limit is
\bel{eq:so4largel}
\tilde{F}^{(0)\ell}_{0,0}(\xp) \underset{\quad\ell \to \infty}{\longrightarrow}
\sum_{j=0}^{\infty} 2^{j+1} (-\xp^{2})^{j} = \frac{1}{\half+\xp^2}
\ \Rightarrow \ \tFF{1}{\ell}{00} = {1\over(\half+\xp^2)^2 }.
\ee
This result can more easily be obtained by noting that the
wavefunction-squared in the large $\ell$ limit is peaked at $v=1/2$
and becomes
$\delta(v-1/2)$.

For generic $\ell$,
the diagonal
form factor approaches the following form at large $n_1$,
\be\label{eq:so4largendist}
\tilde{F}^{(1)\ell}_{n_{1},n_{1}}(\xp)\underset{\quad n_{1} \to
\infty}{\longrightarrow} \frac{1}{\xp (1+\xp^{2})^{3/2}} \ , \ee
which can be obtained by using the same approximation used in
deriving \Eref{rfnlarge}. For finite $n_1$ there is no actual
divergence at $\xp=0$, where the integrals can be done explicitly:
\bel{tFFatzero}\tFF{1}{\ell}{n_1,n_1}(0)=
\frac{2(2n_1+2\ell+1)}{\ell-1} \ . \ee

The computation in momentum space is not as tractable
as that in position space.  For $\FF{1}{\ell=1}{0,n_2}(\q^2)$, the $v$
integration is done in a simple way. We first use an identity
\[
\q^\nu K_\nu ( \q x ) = \int^{\infty}_0 d \m \,
   \frac{\m^{\nu+1} J_\nu ( \m x )}{\q^2 + \m^2}
\]
to write \eref{eq:mespin0} in the following form
\be\label{eq:mespin0j} F^{(1)\ell}_{n_1,n_2} ( \q^2 ) =
\int_0^\infty d\m\, \frac{\m^2}{\q^2 + \m^2} \frac{L^{2}}{2}
\int^1_0 \frac{d v}{v^2} \, \sqrt{v} J_1 ( \m \sqrt{v} )
\phi^{I}_{\ell,n_1} ( v ) \phi^{I}_{\ell,n_2} ( v ) \ . \ee In the
case of $\ell=1$ and $n_1=0$, we use partial integration and the
recurrence relations of the Bessel function in order, to finally
obtain
\begin{multline}\label{eq:intq2m2}
F^{(1)\ell=1}_{0,n_2} ( \q^2 ) = \int_0^\infty d\m\, \frac{m^2}{\q^2 + \m^2}
\cdot \hat{C}^I_{10} \hat{C}^I_{1n_2} (-1)^{n_2} (n_2+2)(n_2+1)\cdot\\
\left(\frac{2}{\m}\right)^3 \left[-\left(\frac{2}{\m}\right) n_2
J_{2 n_2 +3}(\m) + J_{2 n_2 +4}(\m)\right] \ .
\end{multline}

The $\m$ integration can be done in the following way. Since
\[
J_\nu (z)=\half \left[ H^{(1)}_\nu (z) - e^{i \nu \pi} H^{(1)}_\nu (-z)
\right ]
\]
we can convert~\eref{eq:intq2m2} to the integration along the contour shown in
Fig.~\ref{fig:cont}, and we finally have
\begin{multline} \label{eq:me0nspin0}
F^{(1)\ell=1}_{n_2, 0} ( \q^2 ) = 2
\sqrt{3 ( 2 n_2 + 3 )} {{n_{2}+2}\choose 2} \cdot \\
\left(\frac{2}{q}\right)^{2}
\left[
s^{(1)}_{n_{2}}(\q)
- \left( \frac{2}{\q} \right) \left\{ - 2 n_2 K_{2 n_2 + 3} ( \q ) + \q
   K_{2 n_2 + 4} ( \q ) \right\} \right]
\end{multline}
\[
s^{(1)}_{n_{2}}(\q) =
\sum_{j = 0}^{n_2 + 2} \frac{( - 1 )^{n_2 - j} ( n_2 + j ) ! ( j + 1 )}{(
n_2 - j + 2 ) !} \left( \frac{2}{\q} \right)^{2 j }
\]
In particular, the diagonal form factor for the $( \ell, n_{1,2} )
= ( 1, 0 )$ \be\label{eq:me00spin0} F^{(1)\ell=1}_{0,0} ( \q^2 ) =
\frac{12}{\q^2} \left[1-
  \frac{16}{\q^2}+\frac{192}{\q^4} - {4 K_4 ( \q )}\right] .
\ee
\FIGURE[ht]{
\epsfbox{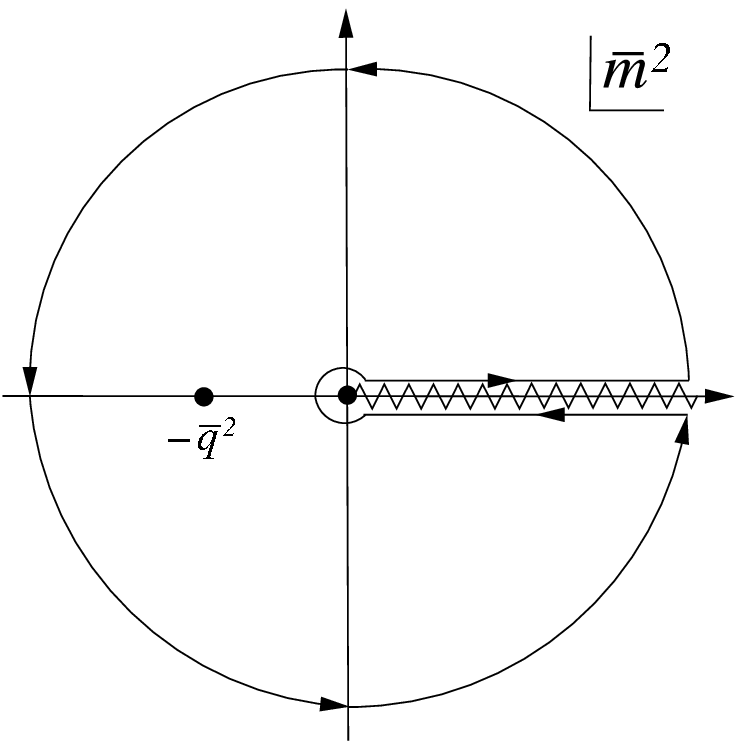}
\caption{The coutour used in the integration of \Eref{eq:intq2m2}.}
\label{fig:cont}
}
Note that $s^{(1)\ell=1}_{n_{2}} ( \q )$ exactly cancels the singular
pieces in the expansion of  the Bessel function around
$\q=0$. This is required in order that $F_{0,n_2}\to \delta_{0,n_2}$
as $q^2\to 0$.
Also note that the form factor goes like $1/q^{2}$ at large $q$, as we
have noticed before.

\rem{
It's interesting to compute examples not involving the lowest lying
state. For the matrix element $\bra{1,1;p}J_4^\mu(q)\ket{1,1;p'}$ and
$\bra{2,0;p}J_4^\mu(q)\ket{2,0;p'}$, the form factors are respectively,
\begin{multline} \label{me11spin0}
F^{(1)\ell=1}_{1,1}(\q^2) = \frac{20}{\q^2} \left[
1- \frac{80}{\q^2} + \frac{6336}{\q^4} - \frac{368640}{q^6}
+ \frac{11796480}{q^8} \right. \\
\left. -\frac{192 ( 160 + 3\q^2 )
     K_3(\q)}{\q^3} -
  \frac{12 ( 20480 + 640\q^2 + 3\q^4 )
     K_4(\q)}{\q^4} \right],
\end{multline}
\begin{multline} \label{me20spin0}
F^{(1)\ell=2}_{0,0}(\q^2) = \frac{640}{\q^4} \left[
1 - \frac{72}{\q^2} + \frac{3456}{\q^4} - \frac{92160}{\q^6}\right.\\
\left. +\frac{3 ( 80 + \q^2 ) K_3(\q)}{\q}
+\frac{48 ( 40 + \q^2 ) K_4(\q)}{\q^2}\right].
\end{multline}
Their Fourier transformations are
\begin{multline}
\tilde{F}^{(1)\ell=1}_{1,1} (\xp)=-\frac{10}6
(23+274\xp^{2}+720\xp^{4}+480\xp^{6})\\
+10
 \frac{d}{d\xp^{2}}\left[
(1+\xp^{2}) \left( P^{2,0}_{1}(2 \xp^{2}-1) \right)^2 \right ]
\log\left(1+\frac1{\xp^{2}}\right) \nonumber
\end{multline}
\begin{multline}
\tilde{F}^{(1)\ell=2}_{0,0}(\xp)= \frac{10}{3}
(3+56\xp^{2}+114\xp^{4}+60\xp^{6})
+40  \frac{d}{d\xp^{2}}\left[\xp^{4} (1+\xp^{2}) \right ]
\log\left(1+\frac1{\xp^{2}}\right)
\end{multline}
We can easily see that these are consistent with our statements about the
general form factors.
}

\subsubsection{Energy-momentum tensor}
We can evaluate the matrix element for the energy-momentum tensor
which couples to the boundary metric. The calculation is similar
to the $SO(4)$ case except  that we deal with the bulk graviton.
We only summarize our results here. The traceless part of the
matrix element is given by
\be\label{eq:empre1}
\bra{\ell,n_{2};p}T^{\mu \nu} \ket{\ell,n_{1};p'}= \left(
\eta^{\mu \rho} - \frac{q^{\mu} q^{\rho}}{q^2} \right) \left(
\eta^{\nu \sigma} - \frac{q^{\nu} q^{\sigma}}{q^2} \right)
\mathcal{T}_{\rho \sigma} \ee \be\label{eq:empre2}
\mathcal{T}_{\rho \sigma} = \left( ( p' + p )_{\rho} ( p' + p
)_{\sigma} - \frac{q^2 ( p' + p )^2 - \{ q \cdot ( p' + p )
\}^2}{3 q^2}\eta_{\rho \sigma}\right) F^{(2)\ell}_{n_{1},n_{2}} (
q^2 ) \ee \be\label{eq:emspin0general}
F^{(2)\ell}_{n_{1},n_{2}}(\q^{2}) = \frac{L^{2}}{4} \int^{1}_{0}
\, \frac{d v}{v^2} \left\{ \q^{2} v K_{2}(\q \sqrt{v})\right\}
\phi^{I}_{\ell,n_1} ( v ) \phi^{I}_{\ell,n_2} ( v ) \ .
\ee

The analysis is parallel to the $SO( 4 )$ case.  In fact, by recurrence
relations between the Bessel functions, we have the relation
\be\label{eq:F4toFg}
F^{(2)} ( \q^2 ) = -\q^4 \frac{d}{d \q^2} \left[ {F^{(1)}(
\q^2 )\over q^2}\right]\ ,
\ee
as follows from relations proven in the appendix.
Hence, the large $q^{2}$ behavior is dominated by the same
leading power as the flavor and the $SO(4)$ cases, $1/q^{2\ell}$.

The Fourier transform of the form factor is also easily carried
out. We obtain
\be\label{eq:frmegrgeneral}
\tilde{F}^{(2)\ell}_{n_{1},n_{2}} ( \xp^2 ) = \frac{L^2}{2} \int d
v \, \frac{4 v^2}{( v + \xp^2)^3} \cdot \phi^{I}_{\ell, n_1} ( v )
\phi^{I}_{\ell, n_2} ( v ) \ , \ee
and the Fourier transform also
has the corresponding relationship with the $S O ( 4 )$ case,
\be\label{eq:FF210} \tilde{F}^{(2)} ( \xp^2 ) =
{1\over\xp^2}\frac{d}{d\xp^2} \left[ \xp^4
   \tilde{F}^{(1)}( \xp^2 )\right] = 2
   \left(\frac{d}{d\xp^{2}}\right)^{2}
\left[\xp^{4} \tilde{F}^{(0)}( \xp^2 )\right] \ee also following
from results given in the appendix. We can follow the same series
of arguments to extract the general behavior. It's the sum of a
rational function in $\xp$ and another rational one multiplied by
$\log(1+1/\xp^2)$. The leading power of the log part in the
small-$\xp$ region is $\xp^{2(\ell-1)}$ again. In the large $\xp$
region, $\tilde{F}(\xp^2) \sim 1/\xp^6$, although the coefficient
of the $1/\xp^6$ tail is non-zero only for $n_1=n_2,n_2\pm
1,n_2\pm 2$.

Because the tail of $\tFF{2}{}{}$ falls now as $1/\xp^6$, the quantity
$\vev{\rgrav^2}$ is finite, unlike $\vev{\rfour^2}$, even
as $\Lambda\to 0$:
\be {\langle \rgrav^2 \rangle} =
\frac{1}{8 m_h^2} \cdot \frac{\ell}{\ell + 1} \left\{ 1 - \frac{
( \ell + 1 )}{6 ( 2 \ell + 3 )} \left( \frac{\Lambda}{m_h}
\right)^2 
\right\} \ .
\ee
For large $\elll$ we have
\bel{rg2largel}
\vev{\rgrav^2}
\underset{\ell\rightarrow\infty}{\longrightarrow} \frac{1}{8
m_h^2}\left\{ 1 - \frac{1}{12} \left( \frac{\Lambda}{m_h} \right)^2
\right\} \ .
\ee
The study of the large $n_1=n_2$ limit for $\ell=1$ gives
\bel{rg2largen}
{\langle \rgrav^2 \rangle}
\underset{n_1\rightarrow\infty}{\longrightarrow}
\frac{1}{8 m_{h}^2}
\left\{ 1 - \frac18 \left( \frac{\Lambda}{m_h} \right)^2 
\right\} \ .
\ee
 From the momentum space point of view, the absence of the
divergence is related to the corresponding
absence of the $\q^{2} \log\q^{2}$ in the
small $\q$ expansion of the non-normalizible mode. Indeed, we have
$\q^{2} v K_{2}(\q \sqrt{v})/2 = 1- \q^{2} v/4 - \q^{4} v^{2} \log
(\q^{2} v)/32 + \cdots$.

Again we may calculate ${\langle \rgrav \rangle}$ for each case that we
studied in the $SO(4)$ context. The actual computation is unnecessary since
we have, from \Eref{eq:FF210},
\be\label{eq:r4torg}
{\langle \rgrav \rangle} =
\frac1{ m_{h}} \int \
\frac{d^2\xp}{2\pi}\,\xp \left[\frac{1}{\xp^2}\frac{d}{d\xp^2}
\xp^{4} \tilde{F}^{(1)}(\xp)\right]
= \frac1{2m_{h}} \int_0^\infty d\xp\, \xp^{2}
\tilde{F}^{(1)} (\xp)  = \frac{\langle \rfour \rangle}{2}.
\ee
Hence, we end the discussion by referring to the results of the previous
section, in particular Eqs.~\eref{eq:vevrfour} and \eref{eq:so4rlargen2}.

\

\noindent\underline{{\bf Examples:}}

\

Since $\tFF{2}{}{}$ can be obtained from $\tFF{1}{}{}$ using \Eref{eq:FF210},
we can use our earlier results on the latter to find, for
instance,
\[
\tilde{F}^{(2)\ell=1}_{0,0} ( \xp ) = - 36(1 + 2 \xp^2) +
6\left[ \left( \frac{d}{d \xp^{2}} \right)^{2} \xp^{4}
(1+\xp^{2})^{2}\right] \log \left( 1 + \frac{1}{\xp^2} \right)
\]
This is consistent with our prevous analysis on the general cases. In
series expansion near $\xp \sim \infty$, we discover that the leading
term is $1/\xp^{6}$ and there is a logarithmic divergence at
$\xp=0$, the latter being absent for $\ell>1$.

In the momentum space, $F^{(2)\ell=1}_{0,n_{2}}(\q^{2})$ has a
similar expression as the $SO(4)$ case,
\begin{multline}
    F^{(2)\ell=1}_{0,n_{2}}(\q^{2})= 2 \sqrt{3 ( 2 n_2 + 3 )} {{ n_2 +
    2 } \choose {2}}\\
\left(\frac{2}{\q}\right)^{2} \left[ s^{(2)}_{n_{2}}(\q)
- \frac{1}{\q} \left\{ \left( 4 n_2 ( n_2 - 1 ) + q^2 \right)
   K_{2 n_2 + 3} ( \q ) + 8 \q K_{2 n_2 + 4} ( \q ) \right\} \right]
   .\nonumber
\end{multline}
\[
s^{(2)}_{n_{2}}(\q) = \sum_{j = 0}^{n_2 + 2} \frac{( - 1 )^{n_2 - j} (
n_2 + j ) ! ( j + 2 ) ( j + 1 )}{( n_2 - j + 2 ) !} \left(
\frac{2}{\q} \right)^{2 j }
\]
In particular, the diagonal matrix element of the lowest lying
mode yields the form factor
\[
F^{(2)\ell=1}_{0,0}(\q^{2})=\frac{24}{\q^2}\left[1- \frac{24}{\q^2} +
\frac{384}{\q^4}
   - \q K_{5}(\q)\right]
\]
Note that the $1/q^{2}$ term, which is leading
at large $q^{2}$, is consistent with our analysis in
general. Near $\q^{2}=0$, it can be easily checked that
$s^{(2)}_{n_{2}}(\q)$ cancels off the all of the singular terms
coming from the Bessel function part. Therefore, the form factor is
regular at $\q^{2}=0$.

\subsection{Spin one hadrons}

We now turn to the spin-one hadrons, type $II$ in Ref.~\cite{Myers},
which are created by acting on the vacuum with the operators $(
Q^{\dag} \Phi^\ell Q - \tilde{Q} \Phi^\ell \tilde{Q}^{\dag} )_{\theta
   \bar{\theta}}$.

\subsubsection{Flavor current}

The coupling of the flavor current to two spin-one hadrons
descends from the following term in the Born-Infeld action.
\begin{equation}
 g_8 \int d^8x \sqrt{-g} g^{\mu\sigma}g^{\nu\kappa}
f^{abc}\partial_{[\mu}A^a_{\nu]}A_{\sigma}^b A_{\kappa}^c.
\label{vertex1}
\end{equation}

In general, the current matrix element $\bra{p', \zeta'} J_{\mu}
\ket{p,\zeta }$ of a spin 1 meson can be arranged into the sum of
electric, magnetic, and quadrupole form factors, 
as in \Eref{vectorformfactors}.
 Plugging the non-normalizable mode for $A_\sigma$ with no
derivative and the normalizable modes for the other $A$'s in
Eq.~(\ref{vertex1}), we get the $F_e$ part of the matrix element. On
the other hand, we get the $F_m$ part of the matrix element by
plugging the non-normalizable mode for $A_{\nu}$ on which
$\partial_{\mu}$ acts and the normalizable modes for the other
$A$'s.   However, there is no quadruple form factor: $F_q=0$.
Moreover, the
electric and  magnetic form factors are equal, $F_e=F_m\equiv F^{(f)}$,
and they are of
the following form:
\[
\FF{f}{\ell}{n_1,n_2}(\q^2) = \frac{R^4}{2} \int_0^1 dw
\left(\frac{w}{1-w}\right)\phi^{II}_{non} (w;\alpha)
\phi^{II}_{\ell, n_1} (w) \phi^{II}_{\ell,n_2}(w).\]
Since $F_e(\q^2\to0)= 1$, the static magnetic ($\mu$) and quadrupole
($D$) moments of the vector meson have no anomalous component:
\begin{eqnarray*}
\mu & = & 1 + F_m(0) = 2, \\
D & = & \frac{2}{m^2}[F_m(0)-2F_q(0)] = \frac{2}{m^2} .
\end{eqnarray*}

The general form of this form factor is the same as that of the
spin-zero case, Eq.~(\ref{ff}), except that the upper limit of the
summation is now $n_1+n_2+2\ell+1$ with $\ell \geq 0$. Following
the same argument as that of the spin-zero case, we get the large
$\q^2$ behavior of the form factor for the general case. In
analogy to \Eref{eq:Spin0LargeQ}, we find the large $\q^2$ behavior
is given by \be\label{eq:Spin1LargeQ}
F^{(f)\ell}_{n_{1},n_{2}}(\q^{2}) \underset{\quad \q^2 \to
\infty}{\longrightarrow} (-1)^{n_1 +n_2} \frac{1}{2}C^{II}_{\ell
n_1}C^{II}_{\ell n_2}\frac{4^{\ell+2} (\ell+2)!}{(\q^2)^{\ell+2}}
\propto \frac{1}{(\q^2)^{\Delta_1 -1} }, \ee where $\Delta_{1} =
\ell+3$ is the conformal dimension of the spin-$1$ mode.

 As in~\cite{IoffeSmilga}, we can classify the form factor into
three parts according to the polarizations of the incoming and the
outgoing hadron states in the Breit frame (the frame in which the initial and
final hadron have equal and opposite momentum vectors.)  In
particular, when both of the polarizations are longitudinal, the form
factor is
\[
F_{LL}=F_{e}-\frac{q^{2}}{2 m^{2}}F_{m}+\frac{q^{2}}{m^{2}}
\left( 1+\frac{q^{2}}{4 m^{2}} \right) F_{q} \ .
\]
where $m$ is the mass of the hadron.
 From $F_{e}=F_{m}$ and $F_{q}=0$, it follows that $F_{LL} \sim
1/(q^{2})^{\Delta_1-2}$.  Equivalently, $F_{LL} \sim
1/(q^{2})^{\tau_1-1}$, where $\tau_1=\Delta_1-1$ is the lowest twist
among all operators which can create this
spin one hadron.  This is consistent with the parton counting rule
which applies at weak coupling. In this regime, it is expected that
$F_{LL}\sim 1/(q^2)^{p-1}$, where $p$ is the number of valence
partons; but $\tau_1=p$ at weak coupling.

  Meanwhile
$F_{LT}$ and $F_{TT}$, where $T$ stands for ``transversely-polarized''
hadrons,  are expected at weak coupling to
be suppressed each by $1/q$ and $1/q^2$, due to
the breaking of helicity conservation~\cite{BrodskyLepage, CZ}. We
similarly, though trivially, find the same behavior at large $\lambda$:
$F_{TT}=F_e\sim 1/(q^2)^{\tau_1-2}$ and $F_{LT}\sim q F_e \sim
1/(q^2)^{\tau_1-3/2}$.

It is straightforward to compute the Fourier transformation of the
form factor. The result is almost the same form as that of the spin-zero
case, \Eref{FTff}, except that the upper limit of the
summation is now $n_1+n_2+2\ell+1$ with $\ell \geq 0$ and there is
no logarithmic divergence for any $\ell$ as $\xp \rightarrow 0$.

The expression for $c^\ell_{n,0,0}$ is given by
\begin{eqnarray}
c^\ell_{n,0,0} &=&  {C^{II}_{0,n}}^2 {C^{II}_{l,0}}^2
B(\ell+3,n+\ell+2){}_3F_2(-n,-n-1,\ell+3;2,-n-\ell-1;1) \nonumber \\
&=& \left\{ \begin{array}{ll}
            (-1)^{\frac{n}{2}} \phantom{0} \frac{4(2n+3)(n+1)!}{[(n/2)!]^2}
                \frac{2 (2\ell+3)! (2+\ell+n/2)!}{(l+1) (\ell-n/2)! (4+2\ell+n)!}
                & \text{$n$  even}\\ \\
            (-1)^{\frac{n+1}{2}} \phantom{0} \frac{2(n+1)(2n+3)(n+2)!}{[(n/2+1/2)!]^2}
                \frac{(2\ell+3)! (3/2+\ell+n/2)!}{(l+1) (\ell-n/2+1/2)! (4+2\ell+n)!}
                & \text{$n$  odd}
            \end{array}
    \right. \label{cspinone} \\ \nonumber \\
&\rightarrow& \left\{ \begin{array}{ll}
                        (-1)^{n/2} \phantom{0} \frac{4(2n+3)(n+1)!}{2^n [(n/2)!]^2}
                            & \text{$n$ even} \\ \\
                        {\rm Order}(1/\ell) \rightarrow 0
                            & \text{$n$  odd}
                        \end{array}
                ,\quad \ell \rightarrow \infty
              \right. \label{cspinone-large-l}
\end{eqnarray}
In the case of $n_1=n_2 \rightarrow \infty$, we get the same
expression for $c_{n,\infty,\infty}^{\ell}$ as in the spin-zero case,
Eq.~(\ref{cspinzero-infty}).

We know that $\langle \rflav^2 \rangle_{\ell\to\infty}$ as well as
$\langle \rflav \rangle_{\ell\to\infty}$ is the same as that of
spin-zero case because $c^\ell_{n,0,0}$ for $\ell \rightarrow \infty$
approach the same limit.\footnote{This can be understood without
calculating $c_{n,0,0}^{\ell=\infty}$ explicitly. Both
${}_3F_2(-n,-n-1,\ell+1;2,-n-\ell-1;1)$ and
${}_3F_2(-n,-n-1,\ell+3;2,-n-\ell-1;1)$ approach the same limit
${}_2F_1(-n,-n-1;2;-1)$ as $\ell \rightarrow \infty$. The ratio of
Beta functions together with the normalization constants also
becomes $1$ as $\ell \rightarrow \infty$.} For any
$\ell$,
\[
m_h \langle \rflav \rangle_{\ell} = \frac{\pi}{2}\sum_{n=0}^{2\ell
+ 1} \frac{c_{n,0,0}^{\ell}}{(\m_n)^3} \qquad,
\]
which has the same behavior with respect to $\ell$ as in the
spin-zero case. $\langle \rflav
\rangle_{n\to\infty}$ is also the same as in the spin-zero case.

 The mean squared radius of the ground state
($n=0$) vector meson of a general $\ell$ is
\be {\langle \rflav^2 \rangle}_{\ell}  =  \frac{1}{m_{h}^2}
\frac{\ell+2}{\ell+1}\left[\psi(2\ell+4)-\psi(\ell+3)\right]
 =  \frac{1}{m_{h}^2} \frac{\ell+2}{\ell+1}
\left[H_{2\ell+3}-H_{\ell+2}\right]. \ee This differs slightly
from \Eref{rfground} but has the same general behavior, and it has
the same large--$\ell$ and large--$n$ limits as
Eqs.~\eref{rfllarge} and \eref{rfnlarge}.
\\

\noindent\underline{{\bf Examples:}}

\

In the case of $\bra{0,0} J_f^{\mu}(q) \ket{0,0}$,
\[\FF{f}{\ell=0}{00}( \q^2 ) =
\frac{12}{\q^2+\m_{0}^2}-\frac{12}{\q^2+\m_{1}^2} \sim
\frac{192}{\q^4}
\]

\begin{eqnarray*}
\tFF{f}{\ell=0}{00}(\xp)
& = & 12 m_h^2 \left[K_0(m_{0} \xp) -K_0(m_{1} \xp) \right]\\
& \rightarrow & 12 m_h^2\sqrt{\frac{\pi}{2m_0\xp}} \quad e^{-m_0 \xp}, \qquad \xp \rightarrow \infty \\
& \rightarrow & 6 m_h^2 \log{3}, \qquad \xp \rightarrow 0
\end{eqnarray*}
Since there is no $1/\q^2$ behavior for large $\q^2$,
$\tFF{f}{\ell=0}{00}(\xp^2) \rightarrow \mbox{constant}$
as $\xp\to 0$.

\rem{
For $\ket{1,0}$,
\[ \FF{f}{\ell=1}{00}( \q^2 ) =
\frac{12}{\q^2+\m_{0}^2}-\frac{60/7}{\q^2+\m_{1}^2}-\frac{12}{\q^2+\m_{2}^2}+\frac{60/7}{\q^2+\m_{3}^2}
\sim \frac{23040}{\q^6}
\]

\[ \tFF{f}{\ell=1}{00}(\xp^2) = m_h^2 \left[12 K_0(m_{0} \xp) -
(60/7)K_0 (m_{1} \xp) -12 K_0(m_{2} \xp)+ (60/7)K_0 (m_{3}\xp)
\right]
\]

\begin{eqnarray*}
 \tFF{f}{\ell=1}{00}(\xp^2)& \rightarrow & 12 m_h^2
\sqrt{\frac{\pi}{2m_0
\xp}} \quad e^{-m_0 \xp}, \qquad \xp \rightarrow \infty \\
& \rightarrow & m_h^2 [6 \ln{6} - (30/7) \ln{(10/3)}], \qquad \xp
\rightarrow 0
\end{eqnarray*}
}

More generally,
\bea
\FF{f}{\ell=0}{0,n_2} ( \q^2 )& =&(-1)^{n_2+1}\times \nonumber \\
& &
\sqrt{\frac{24(n_2+1)(n_2+2)}{(2n_2+3)}}\left[\frac{n_2}{\q^2+\m_{n_2-1}^2}
-\frac{2n_2+3}{\q^2+\m_{n_2}^2}+\frac{n_2+3}{\q^2+\m_{n_2+1}^2}\right].
\nonumber
\eea

\subsubsection{$SO(4)$ current}

One part of the matrix element for the SO(4) current descends from the term
\begin{multline}\label{eq:mevec1}
\frac{\kappa}{R}\int \sqrt{g} \hat A^m v^{\alpha} g^{n r} \partial_{[ m} A_{n ]}
  \partial_{\alpha} A_r = \\
  \frac{\kappa}{R} i \mathcal{Q} \int \sqrt{g} \hat A^m g^{n r} \left\{ (
  \partial_{[ m} A_{n ]} )^{\ast}_f ( A_r^{} )_i - ( A_r^{} )^{\ast}_f (
  \partial_{[ m} A_{n ]} )_i \right\} \ .
\end{multline}
Unlike the spin zero case, there is another contribution from
\begin{multline}\label{eq:mevec2}
\frac{\kappa}{2 R}\int \sqrt{g}(
 -\hat \Gamma^\nu_{\alpha \mu} A_\nu \partial^\alpha A^\mu
+ \hat \Gamma^\mu_{\alpha \nu} A_\mu \partial^\alpha A^\nu) = \\
\frac{\kappa}{R} i \mathcal{Q} \int \sqrt{g}
\partial_\nu \hat A_\mu \left \{ (A^\nu)_f^\ast (A^\mu)_i - (A^\nu)_i (A^\mu)_f^\ast
\right \} \ .
\end{multline}
The matrix element is of the form given in
\eref{vectorformfactors}, but the form of \eref{eq:mevec1} and
\eref{eq:mevec2} implies once again $F_e=F_m\equiv F^{(1)}$ and
$F_{q}=0$. The general form is similar to \Eref{eq:mespin0},
\be\label{eq:mespin1} F^{(1)\ell}_{n_{1},n_{2}} ( \q^2 ) =
\frac{R^2}{2} \int^1_0 \frac{d v}{v} \, ( 1 - v ) \q \sqrt{v} K_1
( \q \sqrt{v} ) \phi^{II}_{\ell,n_1} ( v ) \phi^{II}_{\ell,n_2} (
v ) \ . \ee
Following the same argument as the scalar meson case,
we rediscover \Eref{eq:Spin1LargeQ}.

As before, the Fourier transform is obtained by \eref{eq:F0toF1}
and
\begin{eqnarray}
\tilde{F}^{(0)\ell}_{n_{1},n_{2}}(\xp^2) & = & \frac{R^{2}}{2}
\int^{1}_{0} \frac{d v}{v}(1-v) \frac{\phi^{II}_{\ell n_{1}}
\phi^{II}_{\ell n_{1}}}{v+\xp^{2}}
\label{eq:spin1F0}\\
&=&\Pi_{1}(\xp^{2})\log\left(1+\frac1{\xp^{2}}\right)+\Pi_{2}(\xp^{2})\nonumber
\end{eqnarray}
\[
\Pi_{1}(\xp^2) =\left. \frac{R^{2}}{2} \frac{(1-v)\phi^{II}_{\ell
n_{1}} \phi^{II}_{\ell n_{2}}}{v}\right|_{v=-\xp^{2}}
\]
It is easily seen that the general properties of the vector
mesons are the same as
the scalar mesons, though the degree of the polynomials is different.
Less obvious is the fact that both at large $\ell$, for $n_1=n_2=0$,
and at large $n_1=n_2$, for fixed $\ell$, the spin-one form
factors approach the same limits as the spin-zero form factors,
Eqs.~\eref{eq:so4largel} and \eref{eq:so4largendist}.\footnote{\Eref{tFFatzero}
is however replaced with
$\tFF{1}{\ell}{n_1,n_1}(0)= 2(2n_1+2\ell+3))/({\ell+1})$.}

The charge radius $\left\langle \rfour^{2}\right\rangle$ is similar in
form to that found for the spin-zero case; it also has an infrared
divergence in the $\Lambda\to0$ limit.
\be
{\left\langle \rfour^2 \right\rangle} =
\frac{1}{m_h^2} \left\{ \log \left( \frac{m_h}{\Lambda} \right) +
\frac{1}{2} ( H_{2 \ell + 4} - H_{\ell + 2} ) + \frac{1}{4} 
\right\}
\ .
\ee
For large $\ell$ we find
\be
\vev{\rfour^2}\underset{\quad \ell\rightarrow\infty}{\longrightarrow}
 \frac{1}{m_h^2} \left\{ \log \left(\frac{m_h}{\Lambda} \right)
+ \half \log 2 + \frac14
 \right\}
\ . 
\ee
Note the large $\elll$ limit is the same as the spin zero
case, \Eref{eq:so4r2largel}. This follows from the above-mentioned identity
of the form factors in this regime.  Similarly, we find that the large $n_{1}$
limit for generic fixed $\ell$ is identical to that of the spin zero
case, \Eref{eq:so4r2largen}.

Meanwhile, for the $n_{1}=n_{2}=0$ states, we have
\bel{eq:so4r1spin1}
\left\langle \rfour \right\rangle = \frac{\pi}{2 m_{h}}
\frac{\Gamma(\ell+\frac52) \Gamma(2\ell+4)}{\Gamma(\ell+2)
\Gamma(2\ell+\frac92)}\ .
\ee
Again, the identity of the spin-zero and spin-one form factors
at large $\ell$ implies  $\left\langle \rfour\right\rangle
\to (\pi/2 \sqrt{2}) m_{h}^{-1}$, as in the spin zero
case~\Eref{eq:so4r1largel}.  For fixed $\ell$,
the large--$n_{1}$ limit is $1/m_{h}$, as
for the spin zero result~\Eref{eq:so4rlargen2}.

\

\noindent\underline{{\bf Examples:}}

\

Following the same line of computation as in the spin zero case, we
find the identity
\[
\tilde{F}^{(0)\ell}_{0,n_{2}}(\xp^2) = \frac{\hat{C}^{II}_{\ell 0}
\hat{C}^{II}_{\ell n_{2}}}{n_{2}!(\hat{C}^{II}_{\ell+n_{2},0})^{2}}
\left(\frac{d}{d\xp^{2}}\right)^{n_{2}}
\tilde{F}^{(0)\ell+n_{2}}_{0,0}(\xp^2) ,
\]
and a similar expression for the case of the ground states
\[
\tilde{F}^{(0)\ell}_{0,0}(\xp^2) = (\hat{C}^{II}_{\ell 0})^{2} \left[
\hat{\Pi}_{1}(\xp^{2}) \log\left( 1+\frac{1}{\xp^{2}} \right) +
\hat{\Pi}_{2}(\xp^{2})\right],
\]
\[
\hat{\Pi}_{1}(\xp^{2})=(1+\xp^{2})^{\ell+1}(-\xp^{2})^{\ell+1},\quad
\hat{\Pi}_{2}(\xp^{2})=\mathcal{P}_{++}(\xp^{2}).
\]
where $\mathcal{P}_{++}$ is defined in \Eref{eq:Ppmpm}.
In particular,
\begin{multline}
\tilde{F}^{(1)\ell=1}_{0,0} ( \xp^2 )
= 5 ( 1 - 8 \xp^2 - 66\xp^4 - 60\xp^6 ) + 60
\frac{d}{d\xp^{2}}\left[\xp^6 ( 1 + \xp^{2})^{2} \right]
\log \left( 1 + \frac{1}{\xp^2} \right)
\nonumber
\end{multline}
Note that this is the lowest state that the form factor calculation is
meaningful.  Since the spin one hadron is in the $(\ell/2,\ell/2)$
representation, the $\ell=0$ hadrons are neutral under $SO(4)$
transformation.

In the momentum space, we compute
the matrix element $\bra{\ell=1,n_2}J_4^\mu\ket{1,0}$. The form factor is
\begin{multline}
F^{(1)\ell=1}_{0,n_{2}} ( \q^2 ) = 3\sqrt{5(2 {n_2}+5){{n_2+4}\choose 4}}\\
\left(\frac{2}{q}\right)^{6}
\left [ s^{(1)}_{n_{2}}(\q)
-3 R_{2 n_{2}+7}(4/\q^{2}) \left( \frac{2}{\q}\right) K_{2 n_2 + 7}(\q)
+ R_{2 n_2 +8}(4/\q^{2}) K_{2 n_2 + 8} ( \q )
\right]\ ;
\end{multline}
\[
R_{2n_{2}+7}(z)=( n_2 + 2 ) z^2
   + 14 \binom{n_2 + 3}{3} z + 16 ( 2 n_2 + 7 )
   \binom{n_2 + 3}{4}\ ;
\]
\[
R_{2n_{2}+8}(z) = z^2 + 6 \binom{n_2 + 2}{2} z + 48 \binom{n_2 + 3}{4}\ .
\]
Here $s^{(1)}_{n_{2}}(\q)$ is minus the singular part of the rest of the
expression.
In particular when $n_{2}=0$,
\[
\FF{1}{\ell=1}{00}( \q^2 ) = \frac{23040}{\q^6} \left [ 1 - \frac{96}{\q^2} +
   \frac{3840}{\q^4} + \frac{21 \q + \q^3}{4} K_7 ( \q ) - \frac{24 \q^2 +
   \q^4}{48} K_8 ( \q ) \right] .
\]
We have a leading $1/q^{6}$ term at large $q^{2}$, as predicted.

\rem{
Here are two other
examples.
\begin{multline}
F^{(1)\ell=1}_{1,1} ( \q^2 ) = \frac{161280}{\q^6} \left[ 1 - \frac{288}{\q^2} +
\frac{49920}{\q^4} - \frac{5529600}{\q^6} +
\frac{309657600}{\q^8}\right.\\
\ \ \left.- \frac{3870720 + 92160 \q^2 + 624 \q^4 + \q^6}{48 \q^2} K_4 ( \q )
\right.
\\
\left . \ \
+ \frac{3225600 + 103680 \q^2 + 1000 \q^4 + 3 \q^6}{4 \q^6} K_5 ( \q ) \right]
\end{multline}
\begin{multline}
F^{(1)\ell=2}_{0,0} ( \q^2 ) = \frac{5160960}{\q^8} \left[ 1 - \frac{240}{\q^2} +
\frac{28800}{\q^4} - \frac{1612800}{\q^6}\right. \\
\left. - \frac{2880 + 128 \q^2 + \q^4}{16} K_{10} ( \q ) +
\frac{\q ( 3456 + 144 \q^2 + \q^4 )}{384} K_{11} ( \q ) \right] \ .
\end{multline}
Their Fourier transforms are
\begin{multline}
\tilde{F}^{(1)\ell=1}_{1,1} ( \xp^2 ) = 7(1 - 16\xp^2 - 510\xp^4 -
2300\xp^6 - 3480\xp^8 - 1680\xp^{10})\\
+ \frac{140}{3} \frac{d}{d\xp^{2}} \left[ \xp^6 (1+\xp)^{2}
\left\{P^{2,2}_{1}(-2 \xp^{2}-1)\right\}^{2} \right] \log \left(1 + \frac{1}{\xp^2}\right)
\end{multline}
\begin{multline}
\tilde{F}^{(1)\ell=2}_{0,0} ( \xp^2 ) = \frac{14}{3} ( 1 - 6 \xp^2 + 45
\xp^4 + 500\xp^6 + 870 \xp^8 + 420 \xp^{10} ) \\
- 280  \frac{d}{d\xp^{2}} \left[\xp^8 (1+\xp)^{3} \right]
\log\left( 1 +\frac{1}{ \xp^2} \right)
\end{multline}
}

\subsubsection{Energy-momentum tensor for the spin one case}

We can also compute the energy-momentum tensor matrix element for the
spin one hadrons.  It comes with the similar tensor structure as
\eref{eq:empre1} and \eref{eq:empre2}, and the form factor can be
written in the same way as \eref{eq:emspin0general} except that the
metric factor is now different.  As before, we can derive these form
factors from the $SO(4)$ form factors using Eqs.~\eref{eq:F4toFg} and
\eref{eq:FF210}.

The general properties are similar to the spin zero case.  The form
factor has the same large $q^{2}$ behavior as the $SO(4)$ current,
$1/q^{2(\ell+1)}$. For $n=0$ and general $\ell$,
\begin{eqnarray}
    \left\langle \rgrav^{2} \right\rangle & = &
    \frac{1}{8m_{h}^{2}} \left\{ 1 - \frac{(\ell+3)}{6(2\ell+5)}
    \left(\frac{\Lambda}{m_{h}}\right)^{2} 
\right\} \nonumber \\
    &\underset{\quad \ell \to \infty}{\longrightarrow}&
    \frac{1}{8 m_{h}^{2} } \left\{ 1 - \frac{1}{12}
    \left(\frac{\Lambda}{m_{h}}\right)^{2} 
\right\} \nonumber
\end{eqnarray}
There is no logarithmic divergence.  Again, since the spin-one and
spin-zero form factors are the same in the $\ell\to\infty$ limit ($n_1=n_2=0)$
and in the large $n_1=n_2$ limit for fixed $\ell$, our results for
$\vev{\rgrav^2}$ agree with the spin-zero case in
these computations, namely Eqs.~\eref{rg2largel} and~\eref{rg2largen}.
Moreover, it is
again true from \Eref{eq:r4torg} that
$\langle \rgrav\rangle = \langle \rfour\rangle /2$, and so for $n_1=n_2$
and general $\ell$, $\vev\rgrav$ can be obtained from \Eref{eq:so4r1spin1}.
The large $\ell$ and large $n_1=n_2$ limits can be similarly
read off from
Eqs.~\eref{eq:so4r1largel} and \eref{eq:so4rlargen2}.
\\

\noindent\underline{{\bf Examples:}}

\

Using Eqs.~\eref{eq:F4toFg} and
\eref{eq:FF210}, we find

\[
F^{(2)\ell=0}_{0,0} ( \q^2 ) = \frac{576}{\q^4} \left[ 1 -
\frac{32}{\q^2} - \frac{\q^3}{8} K_5 ( \q ) + \frac{\q^4}{48} K_6 ( \q
) \right]
\]
\[
F^{(2)\ell=1}_{0,0} ( \q^2 ) = \frac{92160}{\q^6} \left[ 1 -
\frac{120}{\q^2} + \frac{5760}{\q^4} + \frac{48 \q^2 + \q^4}{24} K_8 (
\q ) - \frac{48 \q^3 + \q^5}{384} K_9 ( \q ) \right]
\]
and their Fourier transforms are
\[
\tilde{F}^{(2)\ell=0}_{0,0} ( \xp^2 ) = \frac{12}{1 + \xp^2} (1 + 12
\xp^2 + 12 ( \xp )^4) - 12  \left( \frac{d}{d \xp^{2}}
\right)^{2} \left[\xp^6 ( 1 + \xp^2 ) \right] \log \left( 1 + \frac{1}{\xp^2}
\right),
\]
\[
\tilde{F}^{(1)\ell=1}_{0,0} ( \xp^2 ) = 10 ( 1 - 12 \xp^2 - 150 \xp^4 -
180 \xp^6 ) + 60  \left( \frac{d}{d \xp^{2}} \right)^{2}\left[ \xp^8 (
1 + \xp^2 )^{2} \right] \log \left( 1 + \frac{1}{\xp^2} \right).
\]

\section{Implications}

Combining the results of \cite{Myers} with those of the previous
sections reveals a number of unfamiliar patterns.  Taken together,
they confirm that this is a class of bound states quite unlike any
previously studied.

\subsection{The quarkonium spectrum}

Of course the first surprise involves the mass spectrum itself
\cite{Myers}.  As we noted earlier, we would not normally expect
quarkonium of spin $\leq 1$ to be so much lighter than quarkonium
states with higher spin, and certainly not to have a mass of order
$m_h\sim\lambda^{-1/2} m_Q$.  Also surprising is that the masses of
states with radial quantum number $n$, and with $\elll$ of
the light $\Phi_i$ particles added to the (s)quark and anti(s)quark,
is approximately linear in $(n+\elll)$.  One might wonder if a
constituent quark model might apply to these hadrons, though this would
require quark constituent masses of order $m_h$ (much {\it less} than
the bare quark masses) and constituent masses for the $\Phi_i$ of
order $m_h$ (much {\it greater} than any confinement scale $\Lambda$,
if there is any low-energy confinement at all.)

All of these facts are of course straightforward to understand from
supergravity.  In the supergravity limit,
the scale $m_h$ is the only one which appears in the equations solved
by \cite{Myers}.  Moreover, the $n+\elll$ dependence is at least
partially guaranteed by the extra-dimensional Kaluza-Klein-like
structure.  The quantum number $n$ sets the number of nodes of the
hadron's wave function in the $r$ direction, while $\elll$ sets the
number of nodes on the $S^3$, so linear dependence in $n$ when $n\gg
\elll$ and in $\elll$ when $\elll\gg n$ is natural.  But these arguments
give no insight into how to derive these facts independently from
quantum field theory.

  Similarly, the large binding energy of these states is no surprise
from string theory.  The (s)quarks are strings connecting the D3
branes to the D7 brane, and in the supergravity limit are strings
extending from the D7-brane to the horizon of $AdS$.  They are much
longer, and indeed more massive, than the 7-7 strings, which are
massless on the D7 brane.  But from the field theory point of view,
this is still completely mysterious. The strange nature of this
binding is highlighted by the fact that it is vastly reduced for
quarkonium involving two (s)quarks of significantly different mass, or,
even more remarkably, between (s)quarks whose mass parameters $m_1$
and $m_2$ differ only by a phase.  The masses of such states are of
order $|m_1-m_2|$.  The phases of the mass parameters appear in the
interactions of the field theory, but how they conspire to make deep
binding possible when $m_1=m_2$ is unknown.

\subsection{The size of quarkonium}

Perhaps our most striking new observation is this: the form factors
we computed indicate that these states all have sizes of
order $m_h^{-1} = \sqrt{\lambda}/m_Q$, for all $n$ and $\ell$.

 From the supergravity point of view, this is not that hard to
understand.  Just as the supergravity equations ensure that $m_h$ is
the only scale which can determine the masses of hadrons, so it is the
only scale which can appear in the form factors in the nonconfining
theories (where $\Lambda=0$.) Thus for $\Lambda=0$ it must be that
$F(q^2)$ is really $F(\q^2)$ (recall $\q \equiv q/m_h$) and $\tilde
F(x_\perp)$ is really $\tilde F(\xp)$ (recall $\xp = m_h x_\perp$.)
We therefore should not be surprised that $m_Q$ does not appear in any
of our expressions, since any such appearance would require that
$\alpha'$ appear in the bulk physics, through some string theoretic
effect beyond supergravity.

But $m_h$ is not the only possible scale.  A hadron's size, especially
if it contains at least one $\Phi$ particle, certainly could be
infinite in the $\Lambda\to 0$ limit.  We might have expected that at
least some measure of the charge radius of these hadrons would have
been of order $\Lambda^{-1}$. This does not happen (at least for the
conserved currents.)

What is strange about this result is that it differs greatly from what
we would have expected at small 't Hooft coupling.  Let us consider
pure (s)quarkonium first, and then (s)quarkonium with $\Phi$ particles
added; at small 't Hooft coupling these are very different systems.

\subsubsection{Pure (s)quarkonium}

The pure (s)quarkonium system, in which the (s)quark and anti(s)quark
are in a color-singlet state, is hydrogenic --- or more precisely,
positronium-like --- at small 't Hooft coupling.  At $\Lambda=0$ there
is an exactly Coulombic potential between the (s)quarks, so at small
$\lambda$ the size of (s)quarkonium is certainly $\sim(\lambda m_Q)^{-1}$.
As $\lambda$ increases, the system becomes relativistic (for small
angular momentum), so this estimate breaks down.  We do
know on general grounds that the size
of low-lying states will be $r
\sim f(\lambda)/m_Q$, where $f$ is an unknown function that behaves
as $1/\lambda$ at small $\lambda$.  At large $\lambda$,
$f(\lambda)$ could certainly be of order 1, or even
smaller; we have no preconceived notion from the
field theory point of view of how it should behave.

In principle it might have been the case that the quarkonium bound
state approached in appearance a point-particle, with size much less
than its inverse mass, in the large $\lambda$ limit.  If $f$ continued
to shrink, or even went to a constant, this would have been the case.
However, this does not happen; $f(\lambda)$ reaches a minimum
somewhere around $\lambda\sim 1$ and then begins to grow.
The (s)quarkonium states remain as large as, or
larger than (for highly excited states), their inverse masses.
In this sense they retain the fluffy properties of composite
objects even at large $\lambda$.

\subsubsection{Generalized (s)quarkonium}

The behavior  of the states  containing $\Phi$ particles is  even more
difficult  to understand.   For example,  consider the  case  with one
$\Phi$ added  to $Q$ and  $\tilde Q$.  In  this case the  (s)quark and
anti(s)quark are combined in  the adjoint representation of color.  We
might  naively expect,  therefore,  that  they repel,  as  they do  at
leading  order in $\lambda$;  however the  repulsion is  suppressed at
large $N$, so instead we  should think of them as noninteracting.  The
only interaction between them is induced by the light $\Phi$ particle.
In this  sense, this system is  like a hydrogen  molecule, but without
the repulsion between the protons.

At small 't Hooft coupling it is straightforward to carry out a
Born-Oppenheimer computation of this object.  As is easily seen, the
light $\Phi$ particle has a wave function that spreads out over a
distance scale $L_\Phi\sim 1/\lambda m_\Phi$.  This answer is
nearly independent of the distance between the $Q$ and 
$\tilde Q$; it is manifestly true both when the $Q$ and $\tilde Q$ are
well-separated from one another (in which case the $\Phi$ can
be in one of two patches of size $L_\Phi$, one near $Q$ and 
one near $\tilde Q$) and when
the $Q$ and $\tilde Q$ are placed at the same point in space
(in which case the $\Phi$ has a hydrogenic wave function of
size $\sim L_\Phi$.)

The two heavy particles now move in an effective potential induced by
the fact that both are attracted to the $\Phi$ particle.  It is easy
to show that the average size $L_Q$ of the $Q$ and $\tilde Q$ wave
functions is large compared to $1/\lambda m_Q$ and small compared to
$1/\lambda m_\Phi$, independent of the precise details of the
potential.  In the limit $m_\Phi\to 0$, $m_Q$ and $\lambda$ fixed, both
$L_\Phi$ and $L_Q$ diverge.  This also happens in the limit
$\lambda\to0$ with fixed $Q$ and $\Phi$ masses.  On the other hand, in
the limit $m_Q\to\infty$, $m_\Phi$ and $\lambda$ fixed, $L_\Phi$ is constant
and $L_Q\to 0$.  These results would be reflected in the size of
the hadron as measured by the flavor current (which would see only the
heavy particles) and by the $SO(4)$ current (which would be sensitive
to $\Phi$ as well.)

What we have learned about the large-$\lambda$ regime is very
surprising.  We have found that both the flavor and $SO(4)$ currents
see a hadron whose size is of order $\sqrt\lambda/m_Q$.  This is
parametrically smaller than either current would see in the small
$\lambda$ regime.  Taking interesting limits makes this especially
clear.  For $m_\Phi\to 0$, $m_Q$ and $\lambda$ fixed, the size of the
hadron is fixed from the point of view of both currents.  For
$m_Q\to \infty$, $m_\Phi$ and $\lambda$ fixed, both currents see a hadron
shriking to a point; and this is even true when $m_Q\to\infty$,
$m_\Phi\to 0$, with, say, $m_Q m_\Phi$ fixed.  This is a striking
phenomenon not seen previously in quantum field theory, to our
knowledge.  The light, or even massless, $\Phi$ is somehow trapped by
its interactions with the heavy particles at a size scale of order
$m_Q$.

Comparing this behavior with the small-$\lambda$ regime,
we see that from the point of view of $SO(4)$ (and, to a lesser
extent, flavor) the derivative of the hadron's size with respect
to $\lambda$ is extremely large, and negative, near
$\lambda\sim 1$.  As $m_\Phi\to 0$ and/or $m_Q\to\infty$, the
dependence of the size and shape of the hadron on $\lambda$ is apparently
nonanalytic.

\subsection{The meaning of the divergence of $\vev{r_1^2}$}

Now let us turn to the issue of the logarithmic divergence in
$\vev{r_1^2}$.  It is conventional wisdom that a good measure of size,
given a form factor $F(q^2)$, is $\vev{r^2}\propto d[F(q^2)]/d (q^2)$
at $q^2=0$.
However, this idea relies crucially on the exponential falloff of all
charge distributions in position space.  In this model, the
$SO(4)$-charge distributions in position space fall off only as a
power of radius, because the (s)quarks are coupled to a
conformally-invariant sector that carries $SO(4)$ charge.  No matter
how small the coefficient of $1/\xp^4$ might be in a transverse charge
distribution for the $SO(4)$ current, it always leads to a divergent
$\vev{r^2}$.  This proves not that the hadron is infinite in size but
that $\vev{r^2}$ is a bad measure to use.  Indeed other form factors,
even in the \nfour\ sector, have finite $\vev{r^2}$, as does $\vev{r}$
for $SO(4)$.

More physically, if one simply compares graphs of flavor form factors
and of $SO(4)$ form factors in position space, as in
Figs.~\ref{fig:g1}, \ref{fig:g2} and \ref{fig:g3}, one finds they can
be difficult to distinguish, and are never very different.  Instead
one sees that all the hadrons have a core of order $m_h^{-1}$ in size,
and that their power-law tails are only a small part of the structure
of the hadron.  In particular, for large $\ell$ and $n=0$ the fraction
of $SO(4)$ charge stored inside $|\xp|<2$ is $8/9$; for large $n$ and
fixed $\ell$ the fraction is almost the same, $2/\sqrt{5}$.

\FIGURE[ht]{
\epsfbox{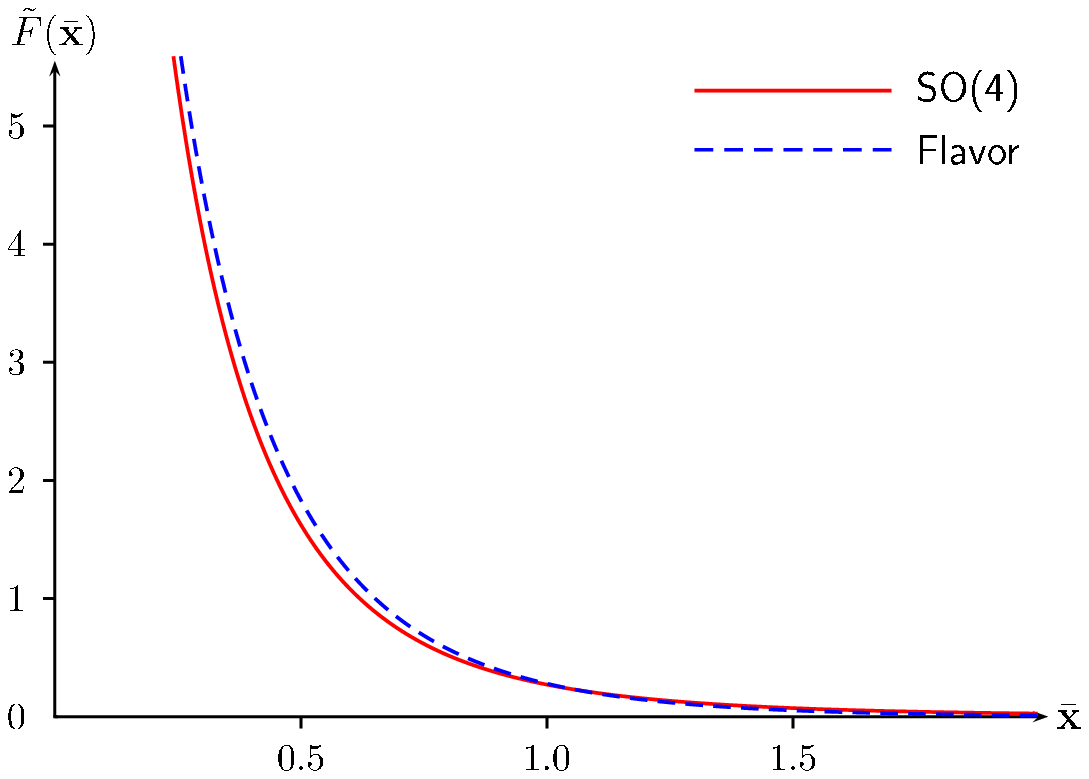}
\caption{The diagonal flavor and $SO(4)$ form factors, in position
space, for the $\ell=1$, $n=0$ spin-zero hadron.}
\label{fig:g1}
}

\FIGURE[ht]{
\epsfbox{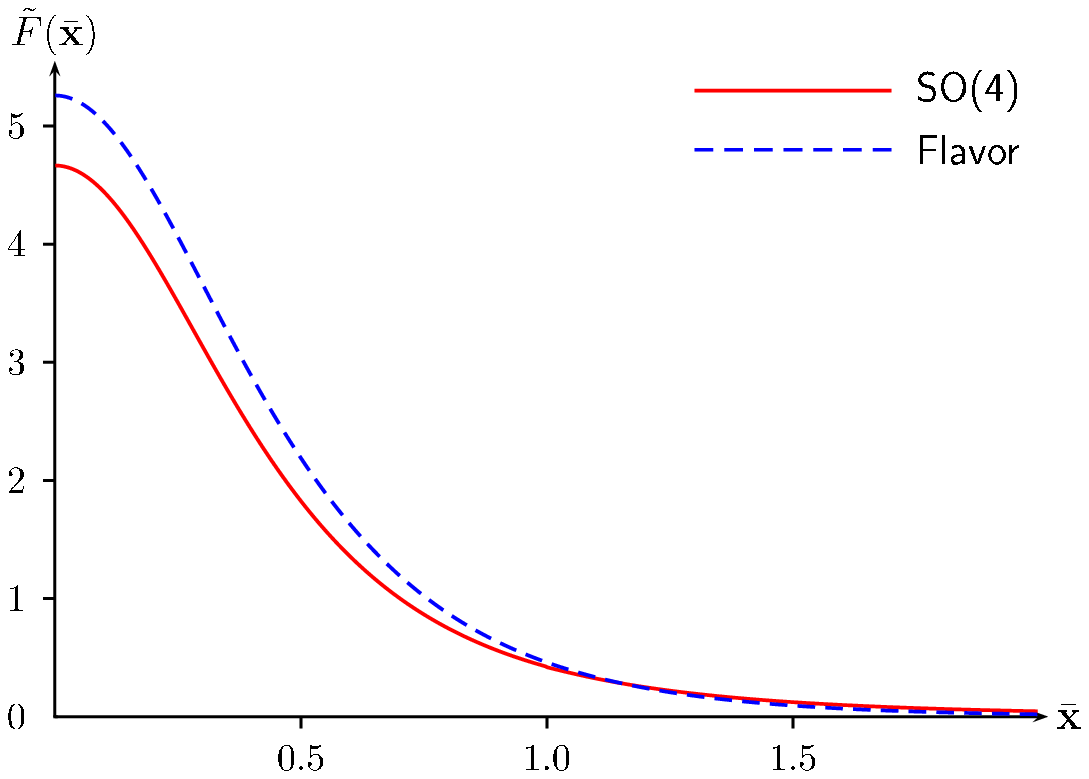}
\caption{The diagonal flavor and $SO(4)$ form factors, in position
space, for the $\ell=10 $, $n=0$ spin-zero hadron.}
\label{fig:g2}
}

\FIGURE[ht]{
\epsfbox{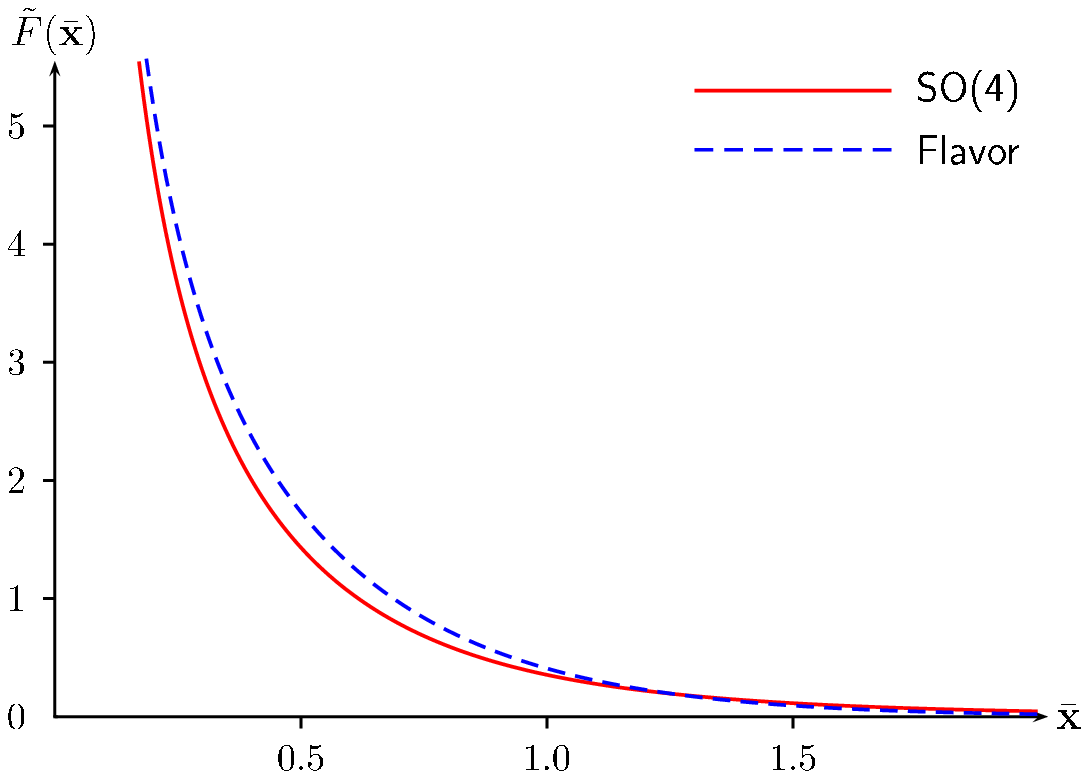}
\caption{The diagonal flavor and $SO(4)$ form factors, in position
space, for the $\ell=1$, $n=10$ spin-zero hadron.}
\label{fig:g3}
}

This fact has implications for string theory.  It has long been said
that strings are infinite in size, because of the divergence found in
$d[F(q^2)]/d(q^2)$ at $q^2=0$ for the energy-momentum form factor.
However, this conclusion might be erroneous.  Strings,
which couple to massless gravitons, {\it etc.}, might well have
charge-distributions and energy-momentum distributions with power-law
tails.  It would be worthwhile to revisit the question of the size of
strings in the light of this observation.

\subsection{The structure of quarkonium}

As we have seen, all of the hadrons in question appear to have a core
and a power-law tail.  The detailed shape of the core, though not its
radial width, depends on $n$ and somewhat on $\ell$.
The tail has a fixed power and a normalization
which is rather insensitive to $n$ and $\ell$.

The flavor form factor tells us where the (s)quarks are located, and
is insensitive to the adjoint matter.  Its properties indicate that
the (s)quarks are located in the core.  The $SO(4)$ current, which is
sensitive to the $\Phi$ fields and gluinos as well as the (s)quarks,
sees a tail for any $\ell$.  That this is true even for $\ell=1$
(spin-zero), for which there are no valence $\Phi$ particles in the
hadron, suggests that the tail is populated by a sea of gluons,
gluinos and $\Phi$ pairs.  (Results from \cite{DIS} suggest that the
sea dominates hadrons at large $\lambda$, in contrast to expectations
at small $\lambda$, where valence particles play a more significant
role.)  The lack of any dramatic difference between the tails of 
$\ell=1$ and $\ell=2$ states suggests that the valence $\Phi$
particles do not strongly change the structure of the hadron.

At large $\ell$ the $SO(4)$ form factor is independent of $\ell$,
while the charge $\QQ$ of the hadron is proportional to $\ell$.
Raising $\ell$ by one corresponds to adding a $\Phi$ particle into the
state; that this has no effect on the shape of the hadron suggests
that for large $\ell$ the valence $\Phi$ particles are all coincident
in position and distribution, affecting the $SO(4)$ charge
distribution only through an overall scaling with $\ell$.

Increasing $n$ does change the shape of the hadron; a somewhat larger
fraction of the flavor charge, $SO(4)$ charge and energy are located
in central part of the core.  However, even in this case
the fractions of the charges located in the tail differs little
between large and small $n$.  It does not seem, then, that
large $n$ has much impact on the basic structure of the state.

In both the large--$\ell$ and large--$n$ limits, in which the hadron
mass also becomes large, the form factors have a simple structure.
The core size is of order $m_h^{-1}$ (which is much larger than the
hadron's inverse Compton length), and the tail has a fixed
normalization and power law.  This suggests that the hadron acts like a
bag; the $\Phi$ particles, trapped inside, have a symmetric wave
function with radius of order $m_h^{-1}$.  The puzzle, as before, is
that the size of this bag is set by $1/m_h$, rather than the
mass-scale of the $\Phi$ particles themselves.

The small size of the bag is directly tied to the deep binding of the
state, in the following sense.  Suppose a $\Phi$ particle's wave
function spreads out to a width $d$, where $\Lambda^{-1}\gg d\gg m_h$.
This arrangement suffers a huge energy cost.  In the stringy dual
description of this phenomenon, this corresponds roughly to a part of
the 7-7 string which represents the hadron descending from the D7
brane, which lies at the $AdS$ radius $L=m_Q\alpha'$, down to an $AdS$
radius of order $r=\sqrt{\lambda}\alpha'/d$, and then returning to the
vicinity of the D7 brane.  This would have an energy cost not of order
$1/m_h$, as one might expect, but of order $2(m_Q-\sqrt{\lambda}/d)$.
Thus the same physics which makes the 7-7 string nearly massless also
provides a mechanism by which it can energetically trap the $\Phi$
particles.  Similarly, if the $\Phi$ wavefunction remains small but
moves a distance $d$ from the center of the bound state, this costs of
order $2\sqrt\lambda m_h^2 d = 2 (m_Q^2/\lambda) d$.  This linear growth
of the energy with $d$ is a less drastic effect than the previous one,
but is still substantial, and is consistent with the Regge-behavior
found in \cite{Myers} for states of moderate spin, which are of linear
size larger than $1/m_h$.    As
emphasized earlier, we still have no idea from the field theory point
of view why the deep binding is present, but it is clear that the
small size and deep binding are correlated.

\subsection{The case of the missing form factors}

Finally we turn our attention to another key observation, which was
first made by Son and Stephanov \cite{SSinprep} in the context of a
recent model \cite{SonSteph}.  These authors noted that there are no
anomalous magnetic form factors or quadrupole form factors for the
spin-one particles in their recent model \cite{SonSteph}.  This is in
contrast to hadrons in QCD, such as the $\rho$ meson
\cite{IoffeSmilga}.  The reason for this can be traced to the
supergravity limit.  The couplings which are required are present in
the string Lagrangian but are subleading in the $\alpha'$ expansion.
The hadrons in our model also share this feature.  For instance,
anomalous flavor magnetic form factors $F_m-F_e$ could only be
generated by a term $\alpha' \tr F_\mu^\nu F_\nu^\sigma F_\sigma^\mu$
in the D7-brane Born-Infeld action.  This term, if present, could only
generate a contribution to $F_m-F_e$ which is suppressed by
$1/\sqrt{\lambda}$.  In a supersymmetric theory, the coefficient of
this term vanishes identically.  But even were supersymmetry broken at
order one, so that the $F^3$ term had a coefficient of order one, the
existence of a low-energy gravity limit would still ensure that the
anomalous magnetic form factor was extremely small.  The quadrupole
form factors are suppressed for the same reasons.  These facts should
have general implications for the structure of hadrons at large 't
Hooft coupling, but we have yet to understand precisely what those
might be.  In particular, despite the fact that these hadrons have a
nontrivial size, there is a sense in which they retain some of the
properties of fundamental point particles in the large $\lambda$
limit.

\subsection{Unanswered questions}

There are many additional computations to consider.  A key issue in
these theories is to understand which aspects of hadronic physics are
determined by scale invariance in the ultraviolet.  Such aspects may
apply approximately in QCD.  We have obtained some results which
probably follow from conformal invariance, possibly in combination
with the large-$\lambda$ limit; it remains to prove this connection
and understand which conditions are necessary and which sufficient.

The shapes and sizes of hadrons in this particular model may or may
not reflect properties of typical light hadrons found in the large
$\lambda$ limit; this remains to be explored.  Our results presumably
can be used to help interpret the structure of light hadrons purely in
the \nfour\ sector.  Previously, the lack of a second scale in
theories such as \nonestar\ has limited our ability both to calculate
and to interpret form factors.  In the present paper, we were able to
separate the hadron mass scale $m_h$ and the confinement scale
$\Lambda$, to great advantage.

On the other hand, the internal structure of these (generalized)
quarkonium states still is unknown.  The methods of \cite{DIS} can and
should be applied to them, in order to better understand what role the
(s)quark and anti(s)quark are playing.  It would be even more
effective to study the full off-forward distribution amplitudes
\cite{muller,ji,radyushkin,burkhart} which can combine these pieces of
information into one package.

Finally, there are a large number of still more challenging
computations to do that go beyond the supergravity approximation.
What happens to these (s)quark-anti(s)quark bound states when one
quark mass is much larger than the other?  or even when their masses
have the same magnitude but different phases, which changes the
interactions and the binding energies drastically?  What is the
structure and the dynamics governing the higher spin states?  Do any
of these stringier states in any way resemble their counterparts in
QCD?  Can we understand, perhaps even quantitatively, how
states with various spins metamorphose from their small-$\lambda$ to
large-$\lambda$ forms?  As of now, the answers to these questions can only
be guessed at, but many of them seem tractable, and there is hope
that considerable progress is possible even with present-day techniques.

\acknowledgments

We thank S. Brodsky,
R. Myers, M. Savage,
M. Stephanov, L. Susskind, L. Yaffe,
and especially D. Son for useful conversations concerning
this subject.  This work was supported by U.S. Department
of Energy grants
DE-FG02-96ER40956 and DOE-FG02-95ER40893,
and by an award from the Alfred P. Sloan Foundation.

\appendix

\section{Decomposition of the Form Factors}

A matrix element $\vev{b,p'|J^\mu(q)|a,p}$ can be written in terms
of a sum over many hadronic states in the supergravity limit. In
particular, the following decomposition of the form factor can be
used:
\bel{Ffg} \quad F_{ab} ( q^2 ) = \sum_n \frac{f_n g_{n
ab}}{q^2 + m_n^2} \ee as illustrated in Fig.~\ref{fig:decomp}.
This decomposition of the form factor into a sum over resonances
is not justified {\it a priori} in a general confining field
theory. However, in the large $N$, large $\lambda$ limit, it is
exact, with corrections of order $1/N$ and $1/\sqrt{\lambda}$.

\FIGURE[ht]{
\epsfbox{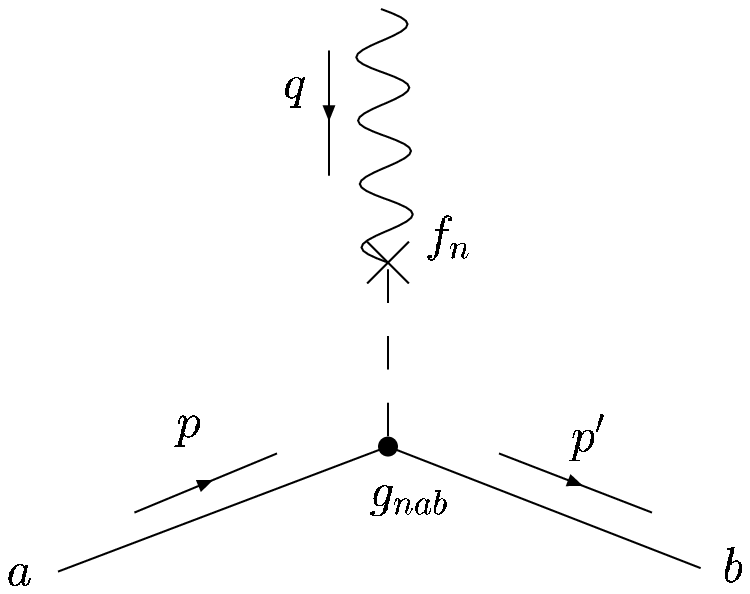}
\caption{Decomposition of the form factor into a sum over hadron
states, as in \Eref{Ffg}.}
\label{fig:decomp}
}

For the spin-one current matrix elements, the $f_n$ are the hadron
decay constants of the spin-one hadrons created by the current
\[ \bra{0} J^{\mu} ( q ) \ket{n, p, \epsilon } = f_n
 \epsilon_\nu \left( \eta^{\mu\nu} - \frac{q^{\mu}q^\nu}{q^2}\right)
\delta^4(p-q) \ ,
\]
where $\ket{0}$ is the vacuum of the thoery, and $\ket{n, p,
\epsilon}$ is the spin-one hadron state with mass $m_n$, momentum
$p$ and polarization $\epsilon_{\mu}$ created by the current
operator $J^{\mu}$. The $f_n$ are associated with the Wronksian
between the normalizable and nonnormalizable mode.\footnote{We
thank D. Son for sharing his derivation of this fact.}  For
instance, consider the modes created by the flavor current, which
satisfy the differential equation for $\phi^{II}(\rho)$ with
$\ell=0$:
\[ \frac{1}{\rho^3}\partial (\rho^3 \partial \phi) +
\frac{m^2}{(1+\rho^2)^2} \phi = 0 \]
This is already written in a self-adjoint form, so the Wronskian
of two solutions is
\be W[\phi_1,\phi_2]= (\phi_1\phi_2^{'} - \phi_2\phi_1^{'}) =
\frac{f}{\rho^3}  ,\ee
 where $f$ is a constant independent of
$\rho$.  The fact that $f=f_n$ when $m=m_n$ follows from the form
of the two-point function of the current. We may evaluate $f_n$ by
evaluating the Wronskian at any value of $\rho$ that is
convenient.  We do so at $\rho\to \infty$, using the fact that the
nonnormalizable mode $\phi_{non}$ associated to a conserved
current goes to a $q^2$-independent constant as $\rho\to\infty$,
while the corresponding normalizable mode $\phi_n$ goes as
$\phi_n(\rho)=\xi_n\rho^{-2} + {\rm order}(\rho^{-3})$.
$$
f_n = \lim_{\rho\to\infty} \rho^3 W[\phi_{non}(m_n),\phi_n] =
\phi_{non}(\rho\to\infty)
\left[\rho^3\frac{\partial\phi_m}{\partial \rho}
\right]_{\rho\to\infty} = -2\phi_{non}^{(0)}\xi_n \
$$
where $\phi_{non}^{(0)}=\phi_{non}(\rho\to\infty)$. For the
flavor-current case,
\[
f_n = (-1)^{n} 2(L^2 /g_8 R^2)C_{0n}^{II}.
\]

 The $g_{nab}$ are the coupling constants between this hadron and the
incoming and outgoing scalar hadrons, which are given in almost
the same way as the matrix element, except that instead of a
nonnormalizable mode of the gauge boson or graviton corresponding
to the current, with arbitrary momentum $q^\nu$, we plug into the
trilinear vertex a {\it normalizable} mode corresponding to the
hadron $n$, on shell. For example, in the decomposition of the
form factor Eq.~(\ref{F}), the corresponding
$g_{n,n_1,n_2}^{\ell}$ is given by
\[ g^\ell_{n, n_1 , n_2} = g_8 \frac{L^2}{2} \int_0^1 dv
\frac{1}{v^{2}} \phi^{II}_{0,n} (v) \phi^{I}_{\ell,n_1} (v)
\phi^{I}_{\ell,n_2} (v) .\]
Combining this result and the previous one with Eqs.~\eref{ff} and \eref{Ffg},
we can now compute
$c^\ell_{n,n_1,n_2}= f_n g^\ell_{n,n_1,n_2}/m_h^2$.

\section{Some Identities}

We begin with proving Eqs.~\eref{eq:F0toF1} and~\eref{eq:FF210}. It's
enough to prove the following identity. Let $\zeta\equiv v+\sigma$
where $\sigma=\xp^2$.  Then for an integer $S\geq 0$,
\bel{vplusxdescent}
{v^S\over \zeta^{S+1}}=
{(-1)^S\over S!}v^S\frac{\partial^S}{\partial \sigma^S}\frac{1}{\zeta}
={1\over S!} \frac{\partial^S}{\partial \sigma^S}
\left(\sigma^S\frac{1}{\zeta}\right)
\ee
where we have used the fact that $v=\zeta-\sigma$ and $S$ derivatives of
$\zeta^{S-n}$, $1\leq n\leq S$, all vanish.

Let's take the two-dimensional Fourier transform
of this with respect to $d^2\xp$, converting it to a function of
$\q^2$, where $\q_\mu= q_\mu/m_h$ is conjugate to $x^\mu$.  The
Fourier transform of $1/\zeta$ is
\[FT\left({1\over
    \zeta}\right) = \frac{1}{2\pi} \int d^2\xp \, e^{i\q\cdot \xp}
{1\over v+\xp^2} = K_0(\q\sqrt{v}).\]
Now we are ready to show that \Eref{Fg01} implies~\eref{Fg02}, and vice versa.
\bea
FT\left({v^S\over \zeta^{S+1}}\right) &=&
{(-1)^S\over S!}v^S\frac{\partial^S}{\partial v^S} FT\left({1\over
    \zeta}\right) = {1\over 2^S S!}(\q\sqrt v)^SK_S(\q\sqrt{v})\nonumber
\\
&=& {(-1)^S \over S!} (\q^2)^S {\partial^S\over \partial(\q^2)^S}
K_0(\q\sqrt{v})\nonumber
\eea

This relation also follows from
\bea
\frac{1}{2\pi}\int d^2 \xp e^{i\q\cdot\xp} {\partial^S\over\partial(\xp^2)^S}
\xp^{2S}\tilde F(\xp^2)
& =&
\frac{1}{2\pi}
\int d|\xp|^2 d\theta \cos(\sqrt{\q^2\xp^2}\cos\theta)
 {\partial^S\over\partial(\xp^2)^S} \xp^{2S}\tilde F(\xp^2)
\cr
&=&
\half
\int d|\xp|^2 J_0(\sqrt{\q^2\xp^2})
 {\partial^S\over\partial(\xp^2)^S} \xp^{2S}\tilde F(\xp^2)
\cr
&=&
(-1)^{S}\half
\int d|\xp|^2 d\theta
\left[ {\partial^S\over\partial(\xp^2)^S} J_0(\sqrt{\q^2\xp^2})
\right] \xp^{2S} \tilde F(\xp^2)
\cr
&=&
(-1)^{S}\half
\int d|\xp|^2 d\theta \left({\q^2\over \xp^2}\right)^S
\left[ {\partial^S\over\partial(\q^2)^S}
J_0(\sqrt{\q^2\xp^2})
\right] \xp^{2S} \tilde F(\xp^2)
\cr
&=&
(-\q^2)^S
{\partial^S\over\partial(\q^2)^S}F(\q^2)\nonumber
\eea
where we have used the fact that the integral is even under
$\cos\theta\to -\cos\theta$, and assumed the function $F(\xp^2)$
is sufficiently convergent at $\xp\to 0$ and $\xp\to\infty$
to allow the integration by parts (which is the case in our
applications of this identity.)

Another useful relation that we used in derivation of \Eref{eq:F00toF0n} is
\be
\frac{\partial^n}{\partial v^n}{v^S\over \zeta^{S+1}}
=
{1\over S!} \frac{\partial^S}{\partial \sigma^S} \sigma^S
\frac{\partial}{\partial v^n}\frac{1}{\zeta}
={1\over S!} \frac{\partial^S}{\partial \sigma^S}
\sigma^S \frac{\partial^n}{\partial \sigma^n} \frac{1}{\zeta}.
\ee

\end{document}